\documentclass[aps,showpacs,graphicx]{revtex4} 
\usepackage{amssymb} 
\usepackage[dvips]{graphicx} 
\usepackage[english]{babel} 
\usepackage{indentfirst} 
\usepackage{amsxtra} 
\usepackage{amsmath} 
\usepackage{supertabular} 
\usepackage{multirow} 
\usepackage [mathcal]{eucal} 
\usepackage {float}  
\usepackage{lscape} 
\usepackage{rotating}

\newcommand{\beq}{\begin{equation}} 
\newcommand{\eeq}{\end{equation}} 
\newcommand{\beqn}{\begin{eqnarray}} 
\newcommand{\eeqn}{\end{eqnarray}} 

\newcommand{\caa}{{\cal A}} 
\newcommand{\cab}{{\cal B}}

\newcommand{\vhfmu}{{v_\mu^{\rm HF}}}  
\newcommand{\uhfmu}{{u_\mu^{\rm HF}}}  
\newcommand{\bcse}{\tilde{\epsilon}}

\newcommand{\dbval}{{R_{B}}} 
\newcommand{\axy}{{\cal W}(\mu \mu';J^\Pi)} 
 
\newcommand{\QRPAF}{QRPA(F)} 

\newcommand{\m}{\mu}
\newcommand{\mI}{\mu'}
\newcommand{\n}{\nu}
\newcommand{\nI}{\nu'}
\newcommand{\sixj}[6]{ \left\{ \begin{array}{ccc} 
                               #1 & #2 & #3 \\ 
                                #4 & #5 & #6  
                               \end{array} 
                        \right\} }

\usepackage[usenames]{color} 
\usepackage{ulem}

\begin{document} 
 
\noindent 
\title{Pairing in spherical nuclei: quasi-particle random phase approximation
calculations with the Gogny interaction
}

\author{V. De Donno} 
\affiliation{Dipartimento di Matematica e Fisica ``E. De Giorgi'', 
  Universit\`a del Salento, I-73100 Lecce, ITALY} 
\author{G. Co'} 
\affiliation{Dipartimento di Matematica e Fisica ``E. De Giorgi'', 
  Universit\`a del Salento, I-73100 Lecce, ITALY and, 
 INFN Sezione di Lecce, Via Arnesano, I-73100 Lecce, ITALY} 
\author{M. Anguiano, A. M. Lallena} 
\affiliation{Departamento de F\'\i sica At\'omica, Molecular y 
  Nuclear, Universidad de Granada, E-18071 Granada, SPAIN} 
\date{\today}

\bigskip 
 
\begin{abstract} 
  We investigate the effects of the pairing in spherical nuclei.  We
  use the same finite-range interaction of Gogny type in the three
  steps of our approach, Hartree-Fock, Bardeen, Cooper and Schrieffer,
  and quasi-particle random phase approximation calculations.  We
  study electric and magnetic dipole, quadrupole and octupole
  excitations in oxygen and calcium isotopes and also in isotones with
  20 neutrons. We investigate the pairing effects on single particle
  energies and occupation probabilities, on the excitation energies,
  $B$-values and collectivity of low-lying states including the
  isoscalar electric dipole and the magnetic dipole excitations, and
  also the giant resonances.  The inclusion of the pairing increases
  the values of the excitation energies in all the cases we have
  studied. In general, the effects of the pairing are too small to
  remarkably improve the agreement with the available experimental
  data.
\end{abstract} 
 
\bigskip 
\bigskip 
\bigskip 

\pacs{21.60.Jz; 25.40.Kv}

\maketitle 

\section{Introduction} 
\label{sec:intro} 
In atomic nuclei, pairing is the most striking phenomenon whose
description requires an extension of the extreme mean-field model. For
example, the evidence that the angular momentum of each even-even
nucleus is zero, without any exception, and that the
angular momentum of each odd-even nucleus coincides with
that of the single particle (s.p.)  level occupied by the unpaired
nucleon can be explained only by assuming the presence between
like-nucleons of an attractive two-body interaction, therefore, by
definition, not described by the one-body mean-field potential.

Pairing phenomena have been described by
extending to the nuclear case the theory of Bardeen, Cooper and
Schrieffer (BCS) originally formulated to study superconductivity in
metals \cite{boh58}.  The BCS theory requires two inputs, the
s.p. energies and wave functions, in our case generated by
Hartree-Fock (HF) calculations, and the pairing interaction. This
theory, usually applied to describe the ground state properties of
even-even open shell nuclei, provides the occupation probabilities of
the s.p. levels, and the quasi-particle energies.  An analogous
description of open shell nuclei is also given by the
Hartree-Fock-Bogolioubov (HFB) theory which is formulated to use a
single input, the effective nucleon-nucleon interaction, to generate
s.p. wave functions, occupation probabilities, and
quasi-particle energies.  

The description of the excited states requires to go beyond the HF+BCS
or the HFB frameworks which are, both, unable to consider collective
effects. For this reason an extension of the Random Phase
Approximation (RPA) theory, called quasi-particle RPA (QRPA)
\cite{bar60}, built to handle pairing and partial occupation
probabilities of the s.p. levels, is commonly used.

In these last few years we have developed a model which treats the
pairing by using a set of s.p. levels generated by a HF calculation
\cite{ang14,ang15,ang16a}. In this approach we use the same
finite-range interaction as nucleon-nucleon force in the HF
calculations and also as a pairing force in the BCS approach.  In
the works quoted above we tested our results against those obtained in HFB
calculations, and we found good agreement between the two
approaches. Specifically, the two types of calculations agree very well for those
quantities which are used in QRPA calculations, i.e. occupation probabilities
and quasi-particle energies.

Encouraged by this result we constructed a QRPA on top of our HF+BCS
approach by using the same finite-range interaction. We have now the
possibility of describing ground and excited states of spherical
open-shell nuclei by using an approach which requires a single input,
the nucleon-nucleon finite-range interaction.  As already pointed out
in the seminal paper of Decharg\'e and Gogny \cite{dec80}, the use of
finite-range interaction provides stability of the pairing results
against the size of the s.p. configuration space.  For this reason,
once the nucleon-nucleon interaction has been chosen, our calculations
do not require any other input parameter related to the physics of the problem.

In the literature one can find various types of QRPA calculations, as
indicated, for example, by the review article of Ref. \cite{paa07}.
Without the ambition of being exhaustive in quoting the works done in
this field, we classify them in three categories. There are
calculations where the s.p. wave functions and
energies are generated by a mean-field potential, for example of
Woods-Saxon or harmonic oscillator type.  The BCS equations are solved
by using a phenomenological pairing interaction whose parameters are
selected to reproduce some properties of the nuclei
investigated. Usually, the same pairing interaction is also used in
QRPA calculations.
In other types of calculations the interaction used in the HF procedure
is different from that used to evaluate the pairing.  These are, for
example, the calculations carried out with the Skyrme interaction
\cite{kha02,yam04,ter05,car12,yuk14} whose zero-range character does not
allow its straightforward use in the pairing sector 
since the results strongly depend on the size of the
quasi-particle configuration space \cite{dec80}.
The last type of calculations use the same
interaction  to generate s.p. levels, pairing and QRPA
excitations.  Our calculations, as well as those of
Refs. \cite{gia03,mar11}, belong to this last category.  
There are also calculations carried out within the relativistic framework that  
can be classified in an analogous way (see for example \cite{paa03,cao05}.)

In the present paper we use our HF+BCS+QRPA approach to
investigate the relevance of the pairing in the different steps of the
calculations. By switching on and off the various terms of our
equations we disentangle the pairing effects in the s.p. energies and
occupation probabilities, from those related to the QRPA approach.

Our article is structured as follows. 
In Sec. \ref{sec:model} we describe HF+BCS and QRPA approach. 
The technical details of the calculations are given in
Sec. \ref{sec:details}. In Sec. \ref{sec:results} we show
the results obtained by using the D1M Gogny interaction
\cite{gor09}. Our calculations have been carried out for a set of
oxygen and calcium isotopes and an isotone chain composed by nuclei
with 20 neutrons.  We discuss results concerning positive and 
negative parity multipoles excitations from the dipole to the
octupole excitations.  We analyze the pairing effects first on the
low-lying excitations and, later, on the giant resonances. The
conclusions of our study are presented in
Sec. \ref{sec:summary}. Finally, in the Appendix
\ref{sec:appoxy} a detailed set of results regarding the $^{18}$O
nucleus is presented, in order to discuss in detail the numerical
convergence of our calculations.  
 
\section{The model} 
\label{sec:model} 

The first step of our approach consists in the description of the
nuclear ground state within the HF+BCS framework. We
first generate a set of s.p. wave functions in a spherical basis by
solving the HF equations. When the iterative method described in
Ref.~\cite{co98b} has reached convergence, we calculate the direct,
Hartree, and exchange, Fock-Dirac, potentials by using the s.p. wave
functions below the Fermi level, and solve the integro-differential HF
equations to generate the s.p. wave functions above it.  Each
s.p. state $|\mu \,m_\mu\rangle \equiv |n_\mu l_\mu j_\mu m_\mu
\rangle$ is characterized by the principal quantum number $n_\mu$, the
orbital angular momentum $l_\mu$, the total angular momentum $j_\mu$,
and its $z$-axis projection $m_\mu$, and has a s.p. energy
$\epsilon_\mu$ that is $2j_\m +1$ times degenerated.

The HF wave functions and energies are used as the starting point to
solve the BCS equations.  From the solution of these equations, we
obtain the Bogolioubov-Valatin $v_\mu^2$ and $u_\mu^2$ coefficients,
which are normalized such that $u_\mu^2+v_\mu^2=1$. The occupation
probability of the $|\mu \rangle$ s.p. state is $v_\mu^2$, while
$u_\mu^2$ is the probability of being empty.  We consider only
spherical nuclei, therefore the BCS ground states are spherically
symmetric and the $v_\m$ and $u_\m$ coefficients are independent of
$m_\m$.

A quantity obtained in the BCS calculations and used in the QRPA is
the quasi-particle energy, defined as
\beq
E_\mu \equiv \sqrt {(\epsilon_\mu-\lambda)^2 + \Delta_\mu^2} 
\, ,
\label{eq:qpenergy}
\eeq
where $\lambda$ is the chemical potential, which is calculated by using the expression
\beq
\lambda \,=\,  \displaystyle \frac{ \displaystyle \sum_\mu \,(2j_\mu +1)\, 
\left( 2v^2_\mu \,+ \, \frac{\epsilon_\mu}{E_\mu} \,-\, 1 \right)}
{\displaystyle \sum_\mu \,(2j_\mu+1)\,\frac{1}{E_\mu} }
\, ,
\label{eq:chempot}
\eeq
and
\beq
\Delta_\mu \, = \, -\, \frac{1}{\sqrt{2j_\mu+1} }\,
\sum_\nu \, \sqrt{2j_\nu+1}\, u _\nu\, v_\nu\, \langle \nu \nu;0| V |\mu  \mu;0 \rangle 
\, 
\label{eq:deltak}
\eeq
where the expression $| \alpha \beta; 0 \rangle$ indicates the
coupling of the $|\alpha \rangle$ and $| \beta \rangle$ s.p. states to
the total angular momentum $J=0$, and $V$ represents the pairing
interaction.  

A QRPA excited state $|k\rangle$ of angular momentum $J$, third
component $M$, parity $\Pi$, and excitation energy $\omega_k$, is
described as a combination of quasi-particle excitations on top of the
ground state $| 0 \rangle$:
\beq
|k\rangle \, \equiv \,|J^{\Pi} M; \omega_k \rangle \, =\, \sum_{\m \leq \mI}
 \left[ X_{\m \mI}^{(k)}(J)\, A^{\dag}_{\m \mI}(JM) 
 \,+\,(-1)^{J+M+1}\,Y_{\m \mI}^{(k)}(J) \, {A} _{\m \mI}(J\, -M)\right]  | 0 \rangle
\, .
\label{eq:qdag} 
\eeq
In the above expression, the condition $\mu\leq \mu'$ prevents the
double counting of the quasi-particle pairs.  To simplify the writing
we did not include the explicit dependence on the parity
$\Pi=(-1)^{l_\m\,+\,l_{\mu'}}$.  The QRPA amplitudes $X$ and $Y$ must
verify the relation \cite{rin80,suh07}
\beq
\sum_{\m\leq \mu'} \left\{ [X_{\m \mI}^{(k)}(J)]^* \, X_{\m \mI}^{(k')}(J) \,-\, 
[Y_{\m \mI}^{(k)}(J)]^* \, Y^{(k')}_{\mu \mu'}(J) \right\} \,=\, \delta_{kk'} \, 
\label{eq:XY1}
\eeq
obtained by imposing the orthonormality of the QRPA eigenstates, and
by using the definition of the quasi-particle pair creation and
annihilation operators
\beq
A^{\dag}_{\m \mI}(JM)\, = \, C_{\m \mI}(J) \, \sum_{m_\m, m_{\mI} } \langle j_\m \,m_\m \, j_{\mI} \,m_{\mI}|J\,M  \rangle\, \alpha^\dag_{\m m_\m}\, \alpha^\dag_{\mI m_{\mu'}} 
\eeq
and
\beq
{A} _{\m \mI}(JM)\,=\, C_{\m \mI}(J) \, \sum_{m_\m, m_{\mI} } \langle j_\m \,m_\m \, j_{\mI} \,m_{\mI}|J\,M  \rangle\, 
\alpha_{\m m_\m}\, \alpha_{\mI m_{\mu'}} 
\, ,
\eeq
where
\beq
C_{\m \mI}(J)\,=\, \frac{\sqrt{1\,+\,(-1)^J \delta_{\m \mI}}}{1\,+\, \delta_{\m \mI}} \, .
\eeq

The quasi-particle creation, $\alpha^\dag_{\m m_\m}$, and
annihilation, $\alpha_{\mu m_\m}$, operators are related to the
particle creation, $c^{\dag}_{\m m_\m}$, and annihilation, $c_{\m
  m_\m}$, operators by means of the Bogolioubov-Valatin transformation
\cite{rin80,suh07}
\beq
\alpha^\dag_{\m \,\pm |m_\m |}\,=\,u_\m \, c^\dag_{\m \, \pm |m_\m |}\, \mp \, v_\m \, c_{\m \,\mp |m_\m |} \, .
\eeq

By using standard techniques the QRPA secular equations can be written
as \cite{rin80,suh07}:
\beqn 
\begin{bmatrix} 
\caa(J) & \cab(J)\\
- \cab^*(J) &-\caa^*(J)\\ \end{bmatrix}
\begin{bmatrix} X^{(k)}(J)\\Y^{(k)}(J) \end{bmatrix}=
\omega_k \begin{bmatrix} X^{(k)}(J)\\Y^{(k)}(J) \end{bmatrix} \, .
\label{eq:QRPA}
\eeqn 
The matrix elements of  $\caa$ and $\cab$ are given by the expressions:
\beqn
\caa_{[\m \mI]J,[\n \nI]J}&=&  (E_\m +E_{\mI})\,\delta_{\m \n}\, \delta_{\mI \nI} \label{eq:mata} 
\\ &&\,+\, C_{\m \mI}(J)\,C_{\n \nI}(J)\, \Big\{ F(\m \mI \n \nI ;J)\, (u_\m \bar{v}_{\mI} u_\n \bar{v}_{\nI}+\bar{v}\leftrightarrow u) \nonumber\\
&&\hspace*{3.6cm}
-\, (-1)^{j_\n +j_{\nI} -J}\, F(\m \mI \nI \n ;J)\, (u_\m \bar{v}_{\mI} \bar{v}_\n u_{\nI} +\bar{v}\leftrightarrow u)\nonumber\\
&&\hspace*{3.6cm}
+\, G(\m \mI \n \nI; J)\, (u_\m u_{\mI} u_\n u_{\nI} +\bar{v}\leftrightarrow u) 
\Big\} \nonumber
\eeqn
and
\beqn
\cab_{[\m \mI]J,[\n \nI]J}\, =\, C_{\m \mI}(J)\,C_{\n \nI}(J)\,  \Big\{
F(\m \mI \n \nI ;J)\, (u_\m \bar{v}_{\mI} \bar{v}_\n u_{\nI}+\bar{v}\leftrightarrow u) && 
\nonumber\\
&&\hspace*{-6.5cm}
 -\, (-1)^{j_\n +j_{\nI} -J}\, F(\m \mI \nI \n ;J)\, (u_\m \bar{v}_{\mI} u_\n \bar{v}_{\nI} +\bar{v}\leftrightarrow u) \nonumber\\
&&\hspace*{-6.5cm}
 -\, G(\m \mI \n \nI;J)\, (u_\m u_{\mI} \bar{v}_\n \bar{v}_{\nI} +\bar{v}\leftrightarrow u)
\Big\} \, .
\label{eq:matb}
\eeqn
The secular QRPA equations are independent of the $m$ quantum 
numbers because we are considering spherical nuclei.
In the previous equations, in order to simplify the writing, we have introduced the 
symbol
\beq
\bar{v}_\m \, = \, (-1)^{l_\mu}\, v_{\mu} \, .
\eeq
On the other hand, the $F$ and $G$ functions in Eqs. (\ref{eq:mata}) and (\ref{eq:matb}) contain the matrix elements of the
interaction; their expressions are:
\beq
F(\m \mI \n \nI; J)\,=\, \sum_{K}\, (-1)^{j_\mI+j_\n+K}\,  (2K+1) \, \sixj{j_\m} {j_{\mI}} {J} {j_\n} {j_{\nI}} {K}\, \langle \m \nI ; K |\overline V | \mI  \n ; K \rangle
\label{eq:effe}
\eeq
and 
\beq
G(\m \mI \n \nI;J) \, =\, \langle \m \mI ; J |\overline V | \n  \nI ; J\rangle \, ,
\label{eq:gi}
\eeq
where 
\beq
\langle \m  \nI ;  J |\overline V | \mI  \n ; J \rangle \,=\,
\langle \m \nI ; J | V | \mI  \n ; J \rangle \,
-\, \langle \m \nI; J | V | \n  \mI ; J \rangle 
\eeq
is the antisymmetrized interaction matrix element.

The transition amplitudes induced by an external operator $Q_J^T$ 
are calculated by using the expression:
\beqn
\langle J^\Pi;\omega_k \| Q^T_J \|0\rangle &=& \sum_{\mu\leq\mu'} \, [u_\mu \, \bar{v}_{\mu'}\,+\,(-1)^J \, \bar{v}_\mu \, u_{\mu'}] \nonumber \\ 
&~&
\hspace*{0.72cm}
\left[X_{\m \mI}^{(k)}(J) \,\langle \mu \| Q^T_J \| \mu' \rangle \,+\,
(-1)^{J+j_\mu-j_{\mu'}}\, Y_{\m \mI}^{(k)}(J) \, \langle \mu' \| Q^T_J \|\mu \rangle \right]\, , \label{eq:tme}
\eeqn
where the double bar indicates the reduced matrix element as
defined by the Wigner-Eckart theorem \cite{edm57}. 

In absence of pairing, Eqs. (\ref{eq:mata}) and
(\ref{eq:matb}) do not reproduce the expressions of the usual RPA
matrices \cite{rin80,suh07}. In this limit, the $F$ terms
(\ref{eq:effe}) are the traditional transition
matrix elements of the RPA. However, the $G$ terms in Eq. (\ref{eq:gi}),
that do not have counterpart in the RPA equations, 
do not vanish, and they are simply decoupled from the
$F$ terms. 

In BCS the number of the particles composing the system is conserved only
on average. The QRPA theory is based on the BCS ground state, which is
not an eigenstate of the particle number, and the corresponding
solutions include terms related to nuclear systems with $A\pm2$
particles.
A clear separation of these spurious components
requires the use of projection techniques on the good number of
particles (see for example \cite{cat94,sch96,mar98}). 

 
\section{Details of the calculations} 
\label{sec:details} 

Our formulation of the HF+BCS+QRPA equations can handle any local
finite-range interaction that may, eventually, include density
dependent terms.  Among the various effective interactions available
in the literature, we have chosen forces of Gogny type \cite{cha07t},
which have been widely used and tested.  We express the effective
interaction as a sum of central, $V_{\rm C}$, 
spin-orbit, $V_{\rm SO}$, and density dependent, $V_{\rm DD}$,
terms:
\beq
V(1,2)\, = \, V_{\rm C}(1,2)\,+\,V_{\rm SO}(1,2)\,+\,V_{\rm DD}(1,2)\, .
\label{eq:vgog}
\eeq
The central term depends on the spin and the isospin of the two
interacting nucleons and has a finite range, the other two terms 
are of zero-range type. 
In HF calculations, in addition to the  force $V$, also the
Coulomb interaction has been considered.  

In the BCS calculations the spin-orbit, $V_{\rm SO}(1,2)$, and the
density-dependent, $V_{\rm DD}(1,2)$, terms do not contribute, the
former one because the interacting pair is coupled to zero angular
momentum, and the latter one by construction in Gogny type
interactions \cite{cha07t}.  In our BCS calculations we use only the
$V_{\rm C}$ term and we do not consider the Coulomb term.  This is the
approach commonly adopted in HFB calculations when Gogny type forces
are used \cite{dec80,ber91,egi95}, and it is justified by the small
effects produced by the Coulomb force.  Specifically, for the nuclei
we investigate in this article, we point out that there is no Coulomb
interaction in the pairing sector for the oxygen and calcium isotopes
where the pairing force is active only between neutrons. For the
$N=20$ isotones we have evaluated the effect of the Coulomb force on
the binding energies and we found relative differences between results
with and without Coulomb forces of few parts on a thousand.

The QRPA calculations have been 
carried out by considering the complete Gogny interaction
(\ref{eq:vgog})
plus the Coulomb force, even if we knew that
the effects of the latter one and of the spin-orbit interaction 
are negligible \cite{don14a}.  

In the next section we present the results obtained by using a
parameterization of the Gogny interaction called D1M \cite{gor09}.  We
have carried out our calculations also with the well known and widely
used D1S parameterization \cite{ber91}. We did not find significant
differences between the results obtained with the two interactions in
what refers to the pairing effects. Therefore, we show only the D1M
results.

We solve the HF equations in $r$-space by imposing bound boundary
conditions at the edge of the integration box
\cite{co98b,bau99,ang11}. In this manner all the s.p. states forming
the configuration space are bound, even those with positive
energy. 
The size of the s.p. configuration space is large enough to
ensure the stability of the BCS results.  The finite-range of
the interactions automatically generates the convergence of the
calculations without additional renormalisation parameters as
it is required, for example, when Skyrme-like interactions are used
\cite{tol11}. In our BCS calculations we have considered all the
s.p. states with energy up to 10 MeV. This configuration space,
together with the Gogny interactions, provides the stability of the
BCS ground-state energy within the keV range.

From the numerical point of view, the critical parameter in the QRPA
calculations is the size of the s.p. configuration space which   
is strictly related to the dimension of the integration box.  
As already pointed out,
in our calculations the continuum part of the s.p. spectrum
is described in terms of bound wave functions with positive
energy. The dimension of the integration box determines 
the s.p. energies and wave functions in the continuum.
This procedure does not consider all the effects
produced by a correct treatment of the continuum part
of the s.p. configuration space. In Ref. \cite{don11a}
we have investigated these effects in RPA calculations. 
The effects in the pairing sector have been studied, for 
example in Ref. \cite{mat05}.

Our experience with continuum RPA calculations indicates that the
observables we have investigated in the present work are 
scarcely sensitive to the exact treatment of the continuum
s.p. spectrum.  In analogy to what we have done in Ref. \cite{don14a},
we have chosen the sizes of the s.p. configuration spaces and the
boxes dimensions by controlling that the centroid energies of the
giant dipole resonances of closed shell nuclei do not change by more
than 0.5 MeV when either the box size or the maximum value of the
s.p. energy are increased. By using the values determined in this
manner we obtain stability of our results.  The number of the
quasi-particle pairs depends on the size of the s.p. space and defines
the dimensions of the QRPA secular matrix (\ref{eq:QRPA}) to be
diagonalized.  

This choice of the configuration space sets below zero the spurious 
$0^+$ excited state due to the breaking of the nucleon number conservation
symmetry \cite{ter05}. The case of the spurious $1^-$ state due to the breaking of the translational
invariance will be discussed in Sec. \ref{sec:ese}.
By using the $^{18}$O nucleus as test example,
we give in Appendix \ref{sec:appoxy} a more detailed presentation of
our method to determine the size of the configuration space.

\section{Results} 
\label{sec:results} 
In this section we present a selection of the results of our
calculations with the aim of studying the role of the pairing force in
the excitation spectrum of spherical open shell nuclei as predicted by
QRPA. Our investigation strategy consisted in switching on and off
the pairing force in the various terms of the equations previously presented, 
therefore three different types of calculations have been carried out.  In
those labelled QRPA all the pairing terms are active, while we have
switched off the $G$ terms in Eqs. (\ref{eq:mata}) and
(\ref{eq:matb}) in those calculations we have called QRPA(F). 
In both type of calculations the s.p. basis is that provided by the HF+BCS
approach. This means that in QRPA(F) the pairing is present
only in the s.p. input where energies, and occupation probabilities
differ from those obtained in a pure HF calculation.
In the third kind of calculation, which we call RPA, the s.p. input is
provided by the HF, and the excited states are
obtained by solving Eqs. (\ref{eq:mata}) and (\ref{eq:matb}) 
without the $G$ terms. In this case, we indicate
as $(\vhfmu)^2$ the occupation probability of the s.p. state
$|\mu\rangle$. Its value is 1 or 0 except for the only partially 
occupied s.p. state in each one of the open-shell nuclei investigated. 

In this article, we present the results obtained for 6
oxygen isotopes, from $A=16$ up to $A=26$, 12
calcium isotopes, from $A=40$ up to $A=62$, and a chain of 
9 isotones with $N=20$, from $^{30}$Ne to $^{46}$Fe. 
All the nuclei we have considered are spherical, magic or 
semi-magic. 

\subsection{Single particle energies and occupation probabilities}

We have already discussed in other publications the effect of
the pairing on the ground-state properties of spherical semi-magic nuclei, 
which we describe with our HF+BCS approach \cite{ang14,ang15,ang16a}.  
Here we address our attention on how the
pairing modifies the occupation numbers and the energies of the
s.p. states that are input of the
QRPA calculations.  
For this purpose we have
considered the quantities
\beq
\delta v_\mu \, = \, \left| v_\mu^2 \, -\,  (\vhfmu)^2 \right| 
\eeq
and
\beq
\delta \epsilon_\mu \, = \,\left| 
             \frac {\bcse_\mu - \, \epsilon_\mu \,} 
                   {\bcse_\mu + \, \epsilon_\mu \,}  \right| \, ,
\label{eq:depsilon}
\eeq
where, in analogy with the HF  s.p. energies $\epsilon_\mu$,
we have introduced the BCS s.p. energies $\bcse_\mu$  
defined as
\beq
\bcse_\mu \, = \, \pm\,E_\mu \, + \, \lambda \, ,
\label{eq:bcse}
\eeq
with $E_\mu$ indicating the quasi-particle energy defined in
Eq. (\ref{eq:qpenergy}).
Here the plus sign is taken for s.p. states where $\epsilon_\mu >
\lambda$, while the minus sign corresponds to those s.p. states with
$\epsilon_\mu < \lambda$. 

\begin{figure}[!b] 
\begin{center} 
\includegraphics [scale=0.4,angle=90]{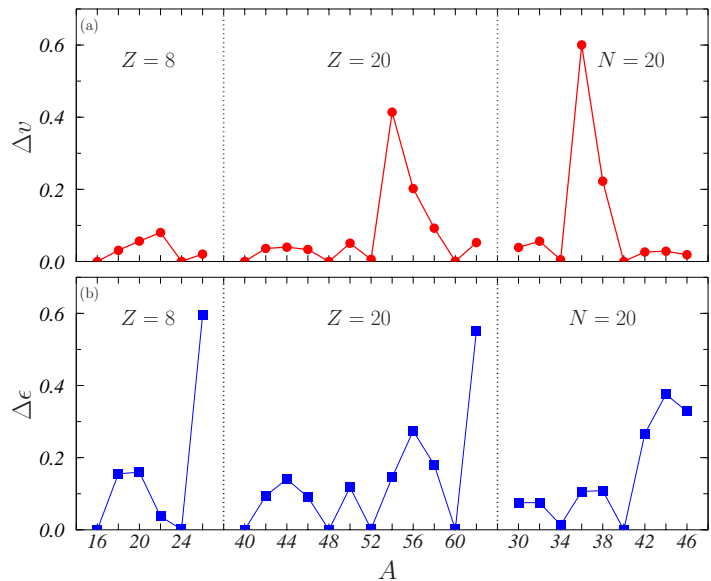} 
\vskip -3.0 mm 
\caption{\small 
Values of (a) $\Delta v$ and (b) $\Delta \epsilon$, defined in Eqs. (\ref{eq:dv2})
and (\ref{eq:deps}), respectively, for the nuclei studied.
  }
\label{fig:ve} 
\end{center} 
\end{figure}

We have found that the values $\delta v_\mu$ and $\delta \epsilon_\mu$
obtained for each individual nucleus are completely
uncorrelated, with Pearson correlation coefficients well below 0.1 in
all cases. The same happens for 
\beq 
\Delta v \,=\,
\sum_\mu \, \delta v_\mu 
\label{eq:dv2}
\eeq
and
\beq
\Delta \epsilon \, = \, 
\sum_\mu \, \delta \epsilon_\mu
\label{eq:deps}
\eeq
calculated for each of the nuclei analyzed.  For these quantities, the
Pearson correlation coefficient is smaller than 0.04.  This scarce
correlation is evident in Fig. \ref{fig:ve} where we show  the values $\Delta v$ and
$\Delta \epsilon$  for all the nuclei we have investigated.  
The zeros in both panels of the figure indicate the closed-shell
nuclei, which are $^{16}$O and $^{24}$O for $Z=8$, $^{40}$Ca,
$^{48}$Ca, $^{52}$Ca, and $^{60}$Ca for $Z=20$ and $^{34}$Si and,
again, $^{40}$Ca for $N=20$.  In these nuclei the pairing is
irrelevant, the s.p. states are fully occupied, or completely empty,
and $\bcse_\mu \sim \epsilon_\mu$.
A clear example of the poor correlation between $\Delta v$ and
$\Delta \epsilon$  is  the case of the $^{22}$O nucleus where  $\Delta
v$ shows the largest value of the isotope chain while that of $\Delta
\epsilon$ is the smallest one of the semi-magic nuclei considered. 
The contrary happens in $^{26}$O.

In panel (a) we notice that the effects of the pairing on $\Delta v$
are similar in all the nuclei studied except in
$^{54}$Ca, $^{56}$Ca, $^{36}$S and $^{38}$Ar where they are notably
larger. This is due to the fact that in these nuclei the HF energy
gaps between the last occupied and the first (partially) unoccupied
s.p.  state are small and this makes the pairing to be more relevant.

In Table \ref{tab:s.p.stat} we present, for these four nuclei, the HF
and BCS occupation probabilities and energies for those s.p. states
with $v_\mu^2> 10^{-3}$.  The table shows that $\bcse_\mu <\epsilon_\mu$
for those quasi-particle states where $\epsilon_\mu < \lambda$, while
the contrary happens if $\epsilon_\mu > \lambda$.  This is a
consequence of the definition of the quasi-particle energy $E_\mu$,
see Eq. (\ref{eq:qpenergy}), where the pairing contribution
$\Delta_\mu$ is present in quadratic form. For this reason, the
inclusion of the pairing always increases the $E_\mu + E_{\mu'}$
factor of the $\caa$ term of the QRPA secular equation (\ref{eq:mata}).

\begin{table}[!t]
\begin{center}
{\scriptsize
\begin{tabular}{c c ccrr  c ccrr c ccrr c ccrr} 
\hline\hline
 &~~~& \multicolumn{4} {c}{$^{36}$S~~~($\lambda=-10.95\,$MeV)} &~~~& \multicolumn{4} {c} {$^{38}$Ar~~~($\lambda=-9.24\,$MeV)} &~~~& \multicolumn{4} {c} {$^{54}$Ca~~~($\lambda=-3.14\,$MeV)}  &~~~& \multicolumn{4} {c} {$^{56}$Ca~~~($\lambda=-2.82\,$MeV)} \\ \cline{3-6} \cline{8-11} \cline{13-16} \cline{18-21}
$\mu$  && $(v^{\rm HF}_\mu)^2$ & $v_\mu^2$ & $\epsilon_\mu$ & $\bcse_\mu$ && $(v^{\rm HF}_\mu)^2$ & $v_\mu^2$ & $\epsilon_\mu$ & $\bcse_\mu$ 
            && $(v^{\rm HF}_\mu)^2$ & $v_\mu^2$ & $\epsilon_\mu$ & $\bcse_\mu$ && $(v^{\rm HF}_\mu)^2$ & $v_\mu^2$ & $\epsilon_\mu$ & $\bcse_\mu$ \\
            \hline
 $1s_{1/2}$ && 1.000 & 0.999 & -44.71 & -44.76 && 1.000 & 0.999 & -42.83 & -42.87 
                   && 1.000 & 1.000 & -49.22 & -49.23 && 1.000 & 1.000 & -48.45 & -48.47 \\
 $1p_{3/2}$ && 1.000 & 0.999 & -30.70 & -30.76 && 1.000 & 0.999 & -29.23 & -29.28 
                    && 1.000 & 1.000 & -36.07 & -36.08 && 1.000 & 1.000 & -35.60 & -35.62 \\
 $1p_{1/2}$ && 1.000 & 0.998 & -27.04 & -27.12 && 1.000 & 0.997 & -25.38 & -25.46 
                    && 1.000 & 1.000 & -32.53 & -32.55 && 1.000 & 0.999 & -32.31 & -32.34 \\
 $1d_{5/2}$ && 1.000 & 0.990 & -17.01 & -17.13 && 1.000 & 0.992 & -16.08 & -16.19 
                    && 1.000 & 1.000 & -22.52 & -22.54 && 1.000 & 0.999 & -22.42 & -22.45  \\
 $2s_{1/2}$ && 1.000 & 0.618 & -11.30 & -12.42 && 1.000 & 0.867 & -10.59 & -11.08 
                    && 1.000 & 0.999 & -17.15 & -17.18 && 1.000 & 0.999 & -17.34 & -17.38  \\
 $1d_{3/2}$ && 0.000 & 0.200 & -10.00 &  -9.37 && 0.500 & 0.570 &  -9.43 & -10.60 
                    && 1.000 & 0.999 & -16.82 & -16.85 && 1.000 & 0.998 & -16.62 & -16.68 \\
 $1f_{7/2}$ && 0.000 & 0.003 &  -2.96 &  -2.90 && 0.000 & 0.005 &  -2.51 &  -2.44 
                    && 1.000 & 0.997 & -10.07 & -10.11 && 1.000 & 0.995 & -10.28 & -10.35 \\
 $2p_{3/2}$ && 0.000 & 0.001 &   0.99 &   1.00 && 0.000 & 0.001 &   1.69 &   1.71 
                    && 1.000 &  0.984 & -5.85 &  -5.94 && 1.000 &  0.985 & -6.02 &  -6.12 \\
 $2p_{1/2}$ &&&&&&&&&&  
                    && 1.000 &  0.717 & -3.46 &  -3.89 && 1.000 & 0.878 &  -3.72 &  -4.00 \\
 $1f_{5/2}$ && 0.000 & 0.001 &   4.35 &   4.37 && 0.000 & 0.001 &   5.15 &   5.17 
                    && 0.000 & 0.104 &  -2.15 &  -1.89 && 0.333 &  0.378 &  -2.56 & -1.73  \\
 $1g_{9/2}$ &&&&&&&&&&  
                    && 0.000 & 0.003 &   1.92 &   1.95 && 0.000 & 0.008 &   1.45 &   1.52 \\
\hline\hline
\end{tabular} 
}
\caption{\small 
Occupation probabilities, $(v^{\rm HF}_\mu)^2$ and $v^2_\mu$, and 
s.p. energies, $\epsilon_\mu$ and $\bcse_\mu$, expressed in MeV, for the four
nuclei showing the largest $\Delta v$ values in Fig. \ref{fig:ve}. 
The s.p. states indicated are proton states for $^{36}$S and
$^{38}$Ar, and neutron states for $^{54}$Ca and $^{56}$Ca.
} 
\label{tab:s.p.stat} 
\end{center} 
\end{table}

The relatively large values of $\Delta \epsilon$  for the  
$^{26}$O and $^{62}$Ca nuclei shown in the panel (b) of 
Fig. \ref{fig:ve}  are generated by the smallness of the denominator
in Eq. (\ref{eq:depsilon}). In these nuclei, the s.p. energies of the 
partially occupied $(1d_{3/2})_\nu$ and  $(1g_{9/2})_\nu$ are close
to 0, in absolute value, in both HF and HF+BCS cases, therefore,
the values of the denominators of the  $\delta \epsilon_\mu$ factors become
comparatively smaller than those of the other nuclei.

\subsection{Excited state energies}
\label{sec:ese}

\begin{figure}[!h] 
\begin{center} 
\includegraphics [scale=0.39,angle=0]{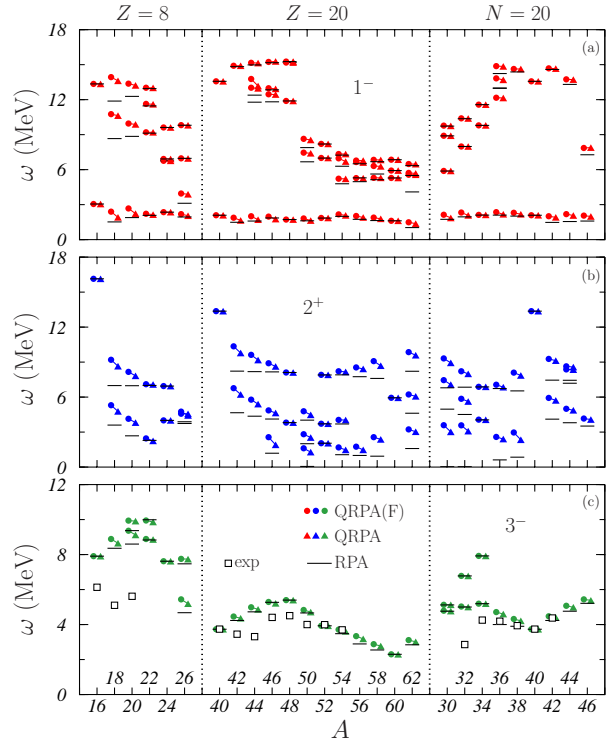} 
\vskip -3.0 mm 
\caption{\small 
Excitation energies of the (a) $1^-$, (b) $2^+$, 
and (c) $3^-$ states for all the nuclei
under investigation.
The black horizontal lines indicate the RPA results, and the circles
and triangles those obtained in QRPA(F) and QRPA calculations, 
respectively. The open squares in panel (c) represent the experimental 
data taken from Ref. \cite{bnlw}.
}
\label{fig:wpar} 
\end{center} 
\end{figure}

After having clarified the effect of the pairing on the input of our
calculations, we present now the QRPA results.  We first discuss the
low-lying excitations. We show in Fig. \ref{fig:wpar} 
the excitation energies, $\omega$, of the main $1^-$, $2^+$ and $3^-$
excited states below the giant resonance region obtained in the RPA
calculations for the nuclei we have considered.
Similar calculations have been carried out for $1^+$, $2^-$ and $3^+$
excitations.  In the figure, the horizontal black lines represent the
RPA results, while the solid circles and the triangles are those
obtained in QRPA(F) and QRPA calculations, respectively. In Fig. \ref{fig:wpar}(c), 
the open squares indicate the experimental energies of  
some $3^-$ excitations taken from Ref. \cite{bnlw}.

In the upper panel of Fig. \ref{fig:wpar}, where the $1^-$ results are
shown, all the nuclei present a first low-lying excitation at about
$2-3\,$MeV. This is a spurious state generated by the breaking
of the translational invariance of the QRPA equations \cite{rin80}.
The presence of this state in the excitation spectrum
is strictly related to the use of a discretized and truncated quasi-particle
configuration space. As we have tested in RPA calculations, which
have analogous problems, a proper treatment of the continuum
s.p. space set to zero this state \cite{don11a}.
For the purposes of the present work, since this state is easily
identifiable and well isolated from the other ones, we have eliminated
it by hand as it has been done in Ref. \cite{co13}, and in the
following discussion is not considered.  
A more detailed discussion about the presence of the spurious 
state is done in the appendix \ref{sec:appoxy}.

\begin{figure}[!b] 
\begin{center} 
\vspace*{-1.2cm}
\includegraphics [scale=0.39,angle=0]{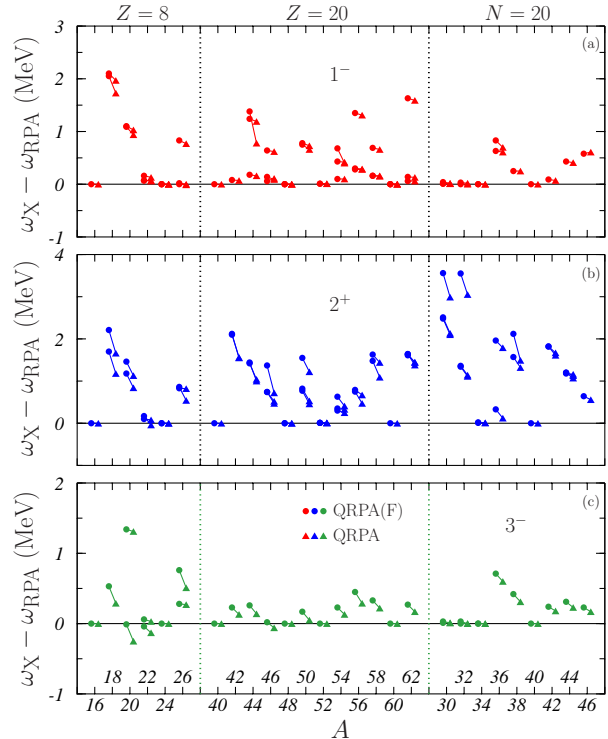} 
\vskip -3.0 mm 
\caption{\small 
Differences between the excitation energies obtained in QRPA(F) 
(circles) and QRPA (triangles) and those found in RPA calculations, 
for the $1^-$, panel (a), $2^+$, panel (b), and $3^-$, panel (c), 
excited states.
}
\label{fig:dwpar} 
\end{center} 
\end{figure}

The first observation about Fig. \ref{fig:wpar} is that the
differences between the results generated by the
various terms containing the pairing force are relatively small as
compared to the values of the corresponding excitation energies. To
have a clearer view of these energy differences we show in
Fig. \ref{fig:dwpar} $\omega_{{\rm QRPA(F)}}-\omega_{\rm RPA}$ (solid
circles) and $\omega_{\rm QRPA}-\omega_{\rm RPA}$ (triangles) for the
same three multipolarities.  It is evident that in the great majority of
the cases considered, the QRPA(F) and QRPA excitation energies 
are larger than the energies obtained in RPA calculations.  
This is, mainly, a consequence of the increase of 
$E_\mu + E_{\mu'}$, previously discussed.

A more compact information about this result is given in Table
\ref{tab:average} where we show, for each of the excitation multipoles
studied, the average values of the differences shown in
Fig. \ref{fig:dwpar}. In this table, also the results obtained for the
unnatural parity excitations $1^+$, $2^-$ and $3^+$ are given.  All
the average values of $\omega_{{\rm QRPA(F)}}-\omega_{\rm RPA}$ and
$\omega_{\rm QRPA}-\omega_{\rm RPA}$ are positive.  As
seen in Fig. \ref{fig:dwpar} we have found maximum differences of
about $2\,$MeV for $1^-$, $3\,$MeV for $2^+$ and $1.5\,$MeV for $3^-$,
while the averages are $1\,$MeV at most.  The table shows that the
values of the standard deviations are rather large, comparable with
the values of the averages.

Since the s.p. input in QRPA and QRPA(F) calculations is the same,
the differences between the energies $\omega_{\rm QRPA}$ and
$\omega_{\rm QRPA(F)}$ are only due to the presence of the
pairing force in the $G$ terms of the QRPA.  The results of
Figs. \ref{fig:wpar} and \ref{fig:dwpar} indicate that, in general,
$\omega_{{\rm QRPA(F)}}$ is larger than $\omega_{\rm QRPA}$.  The
average values of the differences 
$\omega_{\rm QRPA} - \omega_{{\rm QRPA(F)}}$ given in 
Table \ref{tab:average} summarize this
result. These values are all negative and remarkably smaller than the
respective mean of the differences with the RPA energies. The s.p. 
input containing the
pairing, HF+BCS, enhances the excitation energy values, but the
presence of the pairing in the QRPA calculations slightly reduces this
effect.

\newpage \clearpage

\begin{table}[!t] 
\begin{center} 
\begin{tabular}{c c cc c cc c cc} 
\hline  \hline  
   &~~~& \multicolumn{2} {c} {$\omega_{{\rm \QRPAF}}-\omega_{\rm RPA}$} 
   &~~~& \multicolumn{2} {c} {$\omega_{\rm QRPA}-\omega_{\rm RPA}$}  
   &~~~& \multicolumn{2} {c}  {$\omega_{\rm QRPA}-\omega_{{\rm \QRPAF}}$} \\
\cline{3-4} \cline{6-7} \cline{9-10}
 state  && av. & s.d. && av. & s.d. && av. & s.d. \\
\hline
 $1^-$ && 0.35 & 0.39 && 0.31 & 0.35 && 0.00 & 0.20 \\
 $2^+$ && 1.07 & 0.63 && 0.85 & 0.53 &&-0.22 & 0.47 \\
 $3^-$ && 0.21 & 0.21 && 0.14 & 0.19 &&-0.07 & 0.07 \\\hline
 $1^+$ && 0.81 & 0.54 && 0.90 & 0.60 &&-0.08 & 0.58 \\
 $2^-$ && 0.30 & 0.27 && 0.26 & 0.24 &&-0.05 & 0.17 \\
 $3^+$ && 0.72 & 0.54 && 0.69 & 0.52 &&-0.03 & 0.47 \\
\hline\hline  
\end{tabular} 
\caption{\small Averages (av.) and standard deviations (s.d.) of the
  energy differences shown in Fig. \ref{fig:dwpar}, and also of those
  obtained for the $1^+$, $2^-$ and $3^+$  multipole excitations. 
  All the quantities  are expressed in MeV.}
\label{tab:average} 
\end{center} 
\end{table} 

It is also worth remarking that the average energy differences, as
well as the corresponding standard deviations, shown in Table
\ref{tab:average}, are three times larger for positive parity
excitations than for the negative ones. This is due to the fact that,
in the first case, the transitions between spin-orbit partner levels
are allowed. As some of these partner states are close to the Fermi
surface, the pairing produces relatively large effects.

\begin{table}[!b] 
\begin{center} 
\vspace*{0.52cm}
{\scriptsize
\begin{tabular}{c c c cccc c cccc} 
\hline  \hline  
& & & \multicolumn{9}{c}{$\axy$} \\ \cline{4-12}
& & & \multicolumn{4}{c}{$2^+$} &~~&
\multicolumn{4}{c}{$2^-$}\\ \cline{4-7} \cline{9-12} 
 & $A$ &~~& configuration & RPA
& \QRPAF & QRPA &~~& configuration & RPA & \QRPAF & QRPA \\
\hline
$Z=8$ & 16 & && & & & & $(1d_{5/2},1p_{1/2})_{\nu}$ & 0.531 & 0.531 & 0.531 \\
 & 18 & & $(2s_{1/2},1d_{5/2})_{\nu}$ & 0.989 & 0.952 & 0.881 & & $(1d_{5/2},1p_{1/2})_\nu$ & 0.528 & 0.797 & 0.752 \\
 & 20 & &$(2s_{1/2},1d_{5/2})_{\nu}$ & 0.972 & 0.727 & 0.766 & &$(1d_{5/2},1p_{1/2})_{\nu}$ & 0.553 & 0.945 & 0.760 \\
 & 22 & &$(2s_{1/2},1d_{5/2})_{\nu}$ & 0.954 & 0.937 & 0.685 & & & & & \\
 & 24 & &$(1d_{3/2},2s_{1/2})_{\nu}$ & 0.994 & 0.994 & 0.994 & & & & & \\
 & 26 & &$(1d_{3/2},2s_{1/2})_{\nu}$ & 0.921 & 0.937 & 0.772 & &$(2p_{3/2},1d_{3/2})_{\nu}$ & 0.996 & 0.996 & 0.998 \\
$Z=20$ & 40 & & & & & & &$(1f_{7/2},1d_{3/2})_{\nu}$ & 0.520 & 0.520 & 0.520 \\
& 42 &  &$(2p_{3/2},1f_{7/2})_\nu$ & 0.980       & 0.974      & 0.952    & &$(1f_{7/2},1d_{3/2})_\nu$ & 0.567 & 0.768 & 0.736 \\
 & 44 & &$(2p_{3/2},1f_{7/2})_{\nu}$ & 0.979 & 0.944 & 0.921 & &$(1f_{7/2},1d_{3/2})_{\nu}$ & 0.615 & 0.902 & 0.867 \\
 & 46 & &$(2p_{3/2},1f_{7/2})_{\nu}$ & 0.938 & 0.901 & 0.901 & &$(1f_{7/2},1d_{3/2})_{\nu}$ & 0.937 & 0.962 & 0.945 \\
 & 48 & & $(2p_{3/2},1f_{7/2})_{\nu}$ & 0.936 &0.936 &0.936 & & & & & \\
 & 50 & &$(2p_{3/2},1f_{7/2})_{\nu}$ & 0.940 & 0.902 & 0.756 & & & & & \\
 & 50 & &$(2p_{1/2},2p_{3/2})_{\nu}$ & 0.986 & 0.960 & 0.926 & & & & & \\
 & 52 & &$(2p_{1/2},2p_{3/2})_{\nu}$ & 0.984 & 0.984 & 0.978 & & & & & \\
 & 54 & & $(1f_{5/2},2p_{1/2})_{\nu}$& 0.991 & 0.917 & 0.772 & & & & & \\
 & 56 & & $(1f_{5/2},2p_{1/2})_{\nu}$& 0.996 & 0.459 & 0.409 & &$(1g_{9/2},1f_{5/2})_{\nu}$ & 0.974 & 0.929 & 0.889 \\
 & 58 & &$(1f_{5/2},2p_{1/2})_{\nu}$ & 0.999 & 0.899 & 0.785 & &$(1g_{9/2},1f_{5/2})_{\nu}$ & 0.950 & 0.935 & 0.826 \\
 & 60 & & & & & & &$(1g_{9/2},1f_{5/2})_{\nu}$ & 0.936 & 0.936 & 0.936 \\
 & 62 & & & & & & &$(1g_{9/2},1f_{5/2})_{\nu}$ & 0.946 & 0.935 & 0.931 \\
$N=20$ & 30 & &$(2s_{1/2},1d_{5/2})_{\pi}$ & 0.950 & 0.865 & 0.722 & &$(1d_{5/2},1p_{1/2})_{\pi}$ &0.543 & 0.395 & 0.364 \\
 & 32 & & $(2s_{1/2},1d_{5/2})_{\pi}$& 0.938 & 0.907 & 0.868 & &$(1d_{5/2},1p_{1/2})_{\pi}$ & 0.821 & 0.859 & 0.801 \\
 & 34 & &$(2s_{1/2},1d_{5/2})_{\pi}$ & 0.925 & 0.923 & 0.921 & & & & & \\
 & 36 & &$(1d_{3/2},2s_{1/2})_{\pi}$ & 1.003 & 0.903 & 0.827 & & & & & \\
 & 38 & &$(1d_{3/2},2s_{1/2})_{\pi}$ & 0.999 & 0.808 & 0.659 & & $(1f_{7/2},1d_{3/2})_{\pi}$ & 0.520 & 0.824 & 0.806 \\
 & 40 & & & & & & &$(1f_{7/2},1d_{3/2})_{\pi}$ & 0.542 & 0.542 & 0.542 \\
 & 42 & & $(2p_{3/2},1f_{7/2})_{\pi}$& 0.990 & 0.982 & 0.962 & &$(1f_{7/2},1d_{3/2})_{\pi}$ & 0.425 & 0.641 & 0.626 \\
 & 44 & &$(2p_{3/2},1f_{7/2})_{\pi}$ & 0.977 & 0.958 & 0.938 & &$(1f_{7/2},1d_{3/2})_{\pi}$ & 0.420 & 0.831 & 0.806 \\
 & 46 & & $(2p_{3/2},1f_{7/2})_{\pi}$& 
0.966 & 0.926 & 0.922 & &$(1f_{7/2},1d_{3/2})_{\pi}$ & 0.405 & 0.945 & 0.925 \\
 \hline \hline 
 \end{tabular}
 } 
 \caption{\small Values of $\axy$, defined in Eq. (\ref{eq:axy}), 
for the $2^+$ and $2^-$ states obtained in RPA, QRPA(F) and QRPA
calculations. The subscripts $\pi$ and $\nu$ indicate proton
and neutron configurations, respectively. 
}
\label{tab:axy} 
\end{center}
 \end{table}

\newpage \clearpage
\subsection{Collectivity of the excited states}

We have investigated how the pairing modifies the structure of the
wave function of the excited states and, specifically, its
degree of collectivity. To do this we have chosen quasi-particle pair
configurations $(\mu \mu')$ nearby the Fermi level where pairing
effects are more important. In closed shell nuclei we have considered
the pair formed by the HF s.p. states just below and above the Fermi
surface. In open shell nuclei we studied the pairs involving the HF
partially occupied s.p. state and either the first empty or the last fully
occupied level.  Neutron (proton) quasi-particle pair configurations
were selected for $Z=8$ and $Z=20$ ($N=20$) chains.  For each nucleus
and multipolarity $J^\Pi$ we have selected the excited states having
as dominant configuration the quasi-particle pairs chosen and we have
calculated the quantity
\beq
\axy \, = \, 
[X_{\m \mI}(J^\Pi)]^* \, X_{\m \mI}(J^\Pi) \,-\, 
[Y_{\m \mI}(J^\Pi)]^* \, Y_{\mu \mu'}(J^\Pi) \, .
\label{eq:axy}
\eeq
Here $X_{\mu \mu'}$ and $Y_{\mu \mu'}$ are the (Q)RPA amplitudes of
the quasi-particle pair, normalized as indicated in
Eq. (\ref{eq:XY1}). We have followed the evolution of this quantity in
the three types of calculations performed. Obviously, if $\axy \simeq
1$ the excited state is almost a pure quasi-particle pair
configuration, while if the value is remarkably smaller than 1, then,
contributions of other pairs appear and the nuclear excited state is
more collective.
 
The most clear situations are those obtained for the $2^+$ and $2^-$
excitations whose results are shown in Table \ref{tab:axy}. The values
of $\axy$ in RPA results for the $2^+$ excitations are very close to
1, and the inclusion of the pairing diminishes their values. This
indicates that the pairing generates more collectivity. The opposite
effect is present in the $2^-$ excitation. For other multipoles the
situation is more confused and each state has to be investigated
individually.

\subsection{Pigmy dipole resonance}

We have analyzed the low-lying isoscalar $1^-$ excitation known in the
literature as pigmy dipole resonance (PDR). Our results show the
features already identified in RPA \cite{co13} and QRPA \cite{mar11}
calculations with Gogny interactions: the strength of the PDR
increases with the neutron excess. 
On the other hand, the comparison between the results
obtained with the three different type of calculations 
carried out in the present work indicates that
the pairing effects are rather small.

\begin{figure}[ht] 
\begin{center} 
\includegraphics [scale=0.4,angle=0]{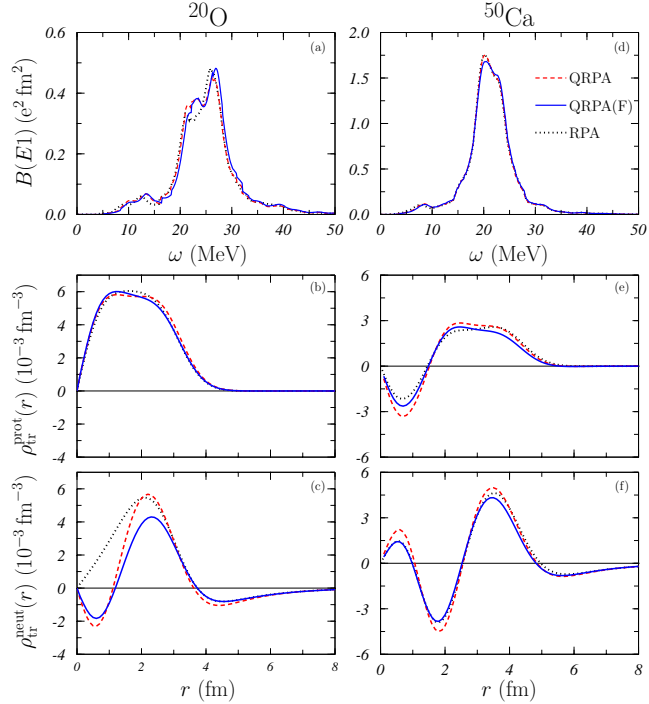} 
\vskip -3mm
\caption{\small Energy distributions of the $B(E1)$ values and
  transition densities for the states identified as PDR in $^{20}$O (left panels)
  and $^{50}$Ca (right panels).  In all panels, red, blue 
  and  black curves indicate QRPA,  QRPA(F) and RPA results, respectively. 
  The $B(E1)$ values energy distributions, shown in panels (a) and (d), 
  have been obtained by folding the results of our calculations with a Lorentz function with
  3 MeV width. Panels (b) and (e) show the proton transition densities, 
  while panels (c) and (f) the neutron ones. In $^{20}$O ($^{50}$Ca) these transition
  densities have been calculated for the states with excitation
  energies of 13.2 (8.6), 13.4 (8.6) and 12.4 (7.9) MeV, for QRPA, QRPA(F) and RPA,
  respectively.}
\label{fig:pdr} 
\end{center} 
\end{figure} 

\begin{figure}[!ht] 
\begin{center} 
\vspace*{-0.7cm}
\includegraphics [scale=0.45,angle=0]{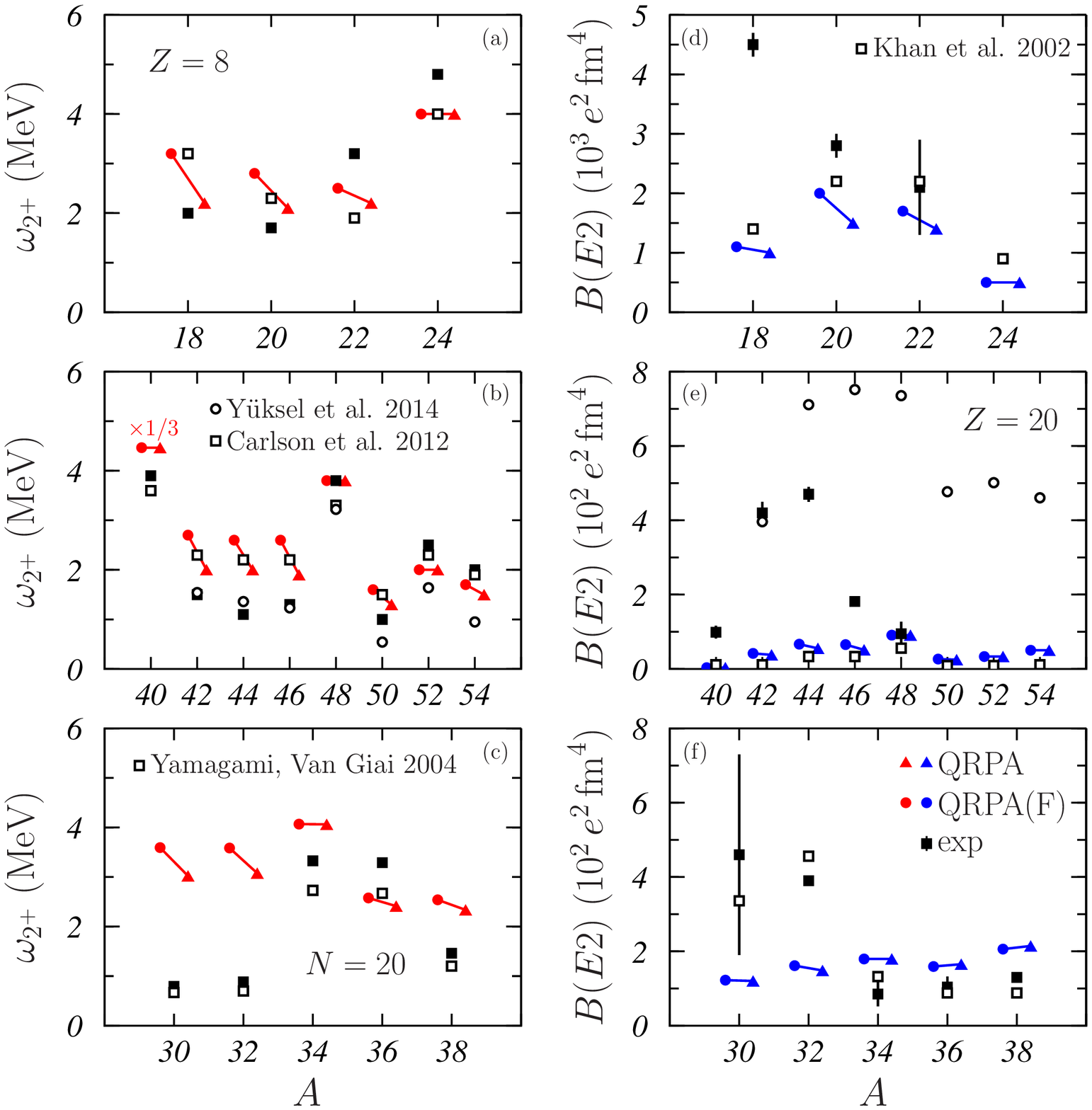} 
\vskip -.3 cm 
\caption{\small Energies (left panels) and $B$-values (right panels) of the lowest 
$2^+$ excitations for some of the nuclei studied with $Z=8$ (panels (a) and (d)), $Z=20$ (panels (b) and (e)) and $N=20$ (panels (c) and (f)). The solid
circles and triangles show our QRPA(F) and QRPA results, respectively. 
The solid squares indicate the experimental data from the 
compilation of Ref. \cite{ram01}, with the exception of 
the $B(E2)$ value of $^{30}$Ne taken from Ref. \cite{yam04}. 
The open squares represent the results of Khan et al. \cite{kha02}, for $Z=8$, 
Carlson et al. \cite{car12}, for $Z=20$, and Yamagami and Van Giai \cite{yam04}, for $N=20$. 
The open circles show the results of Y\"uksel et al. \cite{yuk14} also for $Z=20$. 
Our QRPA and QRPA(F) energy values for the $^{40}$Ca nucleus 
are divided by a factor 3.  
}
\label{fig:other2+} 
\end{center} 
\end{figure} 

In Fig. \ref{fig:pdr} we present the results obtained for the $^{20}$O and
$^{50}$Ca nuclei, where the pairing generates the largest energy
differences between the PDR states.
The energy distributions of the
$B(E1)$ values are shown in the panels (a) and (d), 
the proton transition densities in the panels (b) and (e),
and the neutron transition densities in the panels (c) and (f).
The only remarkable effect of the pairing is in the neutron transition 
density of $^{20}$O, panel (c). In this case, the PDR
excitation is dominated by the $(1f_{7/2},1d_{5/2})_\nu$
quasi-particle pair. We use the subscripts $\pi$ and $\nu$ to indicate proton and neutron configurations, respectively.
In the RPA case there are also remarkable
contributions of other configurations, $(1f_{5/2}, 1d_{5/2})_\nu$,
$(2f_{7/2}, 1d_{5/2})_\nu$ and $(1d_{5/2}, 1p_{3/2})_\nu$, but in QRPA
only the latter configuration contaminates the dominant one. 
In any case, the global effect of the pairing is rather small.

\subsection{Conjugate configurations}

We have already pointed out that in Fig. \ref{fig:wpar} we have
considered those low-lying states which are well identified in RPA and
the corresponding states obtained in QRPA(F) and QRPA calculations.
In both QRPA(F) and QRPA, other low-energy excited states appear
dominated by a quasi-particle configuration implying a s.p. transition
between a partially occupied state and itself. We call {\it conjugate}
these configurations. They play an important role in these
calculations where the pairing interaction is active. However, in RPA
they generate spurious excited states with energy nearby zero, and
give a very small contribution in the wave functions of the other
excited states.
 
In Fig. \ref{fig:other2+} we present energies (left panels) and
$B(E2)$ values (right panels) of the lowest \QRPAF and QRPA $2^+$
states, most of them dominated by conjugate configurations. Our results,
indicated by the solid circles and triangles, are compared to the
experimental data (solid squares) obtained from Ref. \cite{ram01},
with the exception of the $B(E2)$ value of $^{30}$Ne which was
measured more recently \cite{yam04}.
As first general remark, we observe that all the \QRPAF energies are
larger than those obtained in QRPA calculations. This is the same
trend seen in Fig. \ref{fig:wpar}.

In our calculations, the $2^+$ excited states of the oxygen isotopes,
whose energies and $B(E2)$ values are shown in Figs. \ref{fig:other2+}(a) and \ref{fig:other2+}(d), are
all dominated by the conjugate $(1d_{5/2},1d_{5/2})_\nu$
configuration, with the exception of the $^{24}$O nucleus.  The
neutron $1d_{5/2}$ state is completely full in $^{22}$O and $^{24}$O,
but while in the first nucleus the $(2s_{1/2}, 1d_{5/2})_\nu$
configuration competes with the dominant conjugate one, in the second
nucleus the $2^+$ excitation is dominated by the
$(1d_{3/2},2s_{1/2})_\nu$ configuration.
The results of Ref. \cite{gia03} are closer 
to our \QRPAF \,  energies than to those of the QRPA.

In the two panels, the open squares represent the results of the
non-relativistic HFB+QRPA calculations of Ref. \cite{kha02}, which
have been carried out by using the Skyrme interaction SLy4
\cite{cha98} and a zero-range density-dependent interaction pairing
force.  The agreement between our results and those of
Ref. \cite{kha02} is good, certainly more satisfactory than the
description of the experimental data. In this respect we do not
observe any general trend and each case should be separately
discussed.

We found that also the results of the relativistic calculations of
Refs. \cite{paa03,cao05} show behaviors analogous to ours. The
relativistic approach without dynamical pairing, which should
correspond to our QRPA(F), generates excitation energies larger than
those of the RPA. The inclusion of the pairing in the full
relativistic QRPA reduces the values of these energies, as it happens
in our calculations. 

The results regarding the calcium isotopes are shown in the
Figs. \ref{fig:other2+}(b) and \ref{fig:other2+}(e). Up to the $^{48}$Ca isotope, the $2^+$ states are also
dominated by a conjugate configuration, specifically the
$(1f_{7/2},1f_{7/2})_\nu$.  In $^{48}$Ca the neutron $1f_{7/2}$ state
is fully occupied and the excited state is dominated by the $(2p_{3/2}
, 1f_{7/2})_\nu$ pair.  In the case of the heavier $^{50}$Ca isotope,
the $(2p_{3/2} , 2p_{3/2})_\nu$ conjugate configuration gives the main
contribution to the lowest $2^+$ state.  On the other hand, in
$^{52}$Ca, where the $2p_{3/2}$ neutron state is fully occupied, the
excited state is dominated by the $(2p_{1/2} , 2p_{3/2})_\nu$ pair.

In Figs. \ref{fig:other2+}(b) and \ref{fig:other2+}(e) the open squares and circles represent the
results of Refs. \cite{car12} and \cite{yuk14} respectively. The
former have been obtained in HFB+QRPA non-relativistic calculations
which used the SkM* parameterization \cite{bar82} of the Skyrme
interaction and a separable Gaussian interaction as pairing force. The
calculations of Ref.  \cite{yuk14} considered the SLy5 Skyrme
interaction \cite{cha98} and a zero-range density-dependent pairing
force.

Our results are in agreement with those of Ref. \cite{car12} in both
energy and $B(E2)$ values, with the exception of the $^{40}$Ca nucleus
which we shall discuss below. We predict excitation energies larger
than those experimentally found, except for $^{52}$Ca and $^{54}$Ca,
and $B(E2)$ values smaller than the observed ones, except for
$^{48}$Ca.  The calculations of Ref. \cite{yuk14} generate smaller
values of the excitation energies and very large $B(E2)$ values, even
larger than those measured.

The 2$^+$ state of $^{40}$Ca in Ref. \cite{car12} is found at an
energy much lower than that we have obtained 
(see panel (b) of Fig. \ref{fig:other2+}), but
the authors of this article suggest the possibility that this $2^+$ 
is a spurious state with a wrong number of nucleons.  

The results found for the $N=20$ isotones are shown in the Figs. (c)
and (f), where the open squares represent
the values calculated in Ref. \cite{yam04} by using the SkM*
interaction \cite{bar82} plus a zero-range density-dependent
interaction pairing force. The lowest $2^+$ states of the $^{34}$Si,
$^{36}$S and $^{38}$Ar nuclei are dominated by the
$(2s_{1/2} ,1d_{5/2})_\pi$,  the $(1d_{3/2} , 2s_{1/2})_\pi$ and the
conjugate $(1d_{3/2} ,1d_{3/2})_\pi$ pairs, respectively. For these
nuclei, our results describe reasonably well the experimental
data. However, the discrepancies for the $^{30}$Ne and $^{32}$Mg
excitations, dominated by the conjugate $(1d_{5/2}, 1d_{5/2})_\pi$
pair, are remarkable. By using a strong pairing interaction the results
of Ref. \cite{yam04} are able to reproduce the small experimental
energies and the large $B(E2)$ values.

\begin{figure}[ht] 
\begin{center} 
\includegraphics [scale=0.4,angle=90]{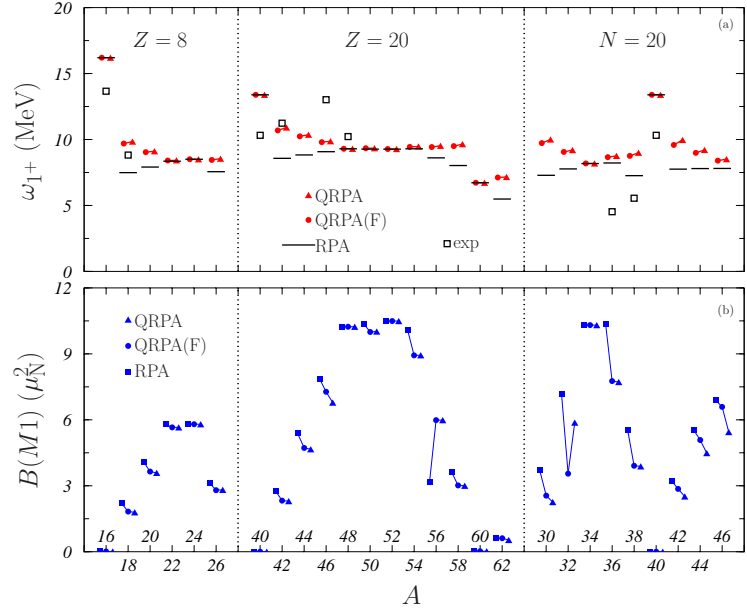} 
\vskip -0.3cm
\caption{\small 
Energies, panel (a), and $B(M1)$ values, panel (b), of 
the $1^+$ states for all the nuclei we have studied.
For each nucleus, we selected the $1^+$ state with
largest $B(M1)$ value. The results of the QRPA(F) and 
QRPA calculations are shown by the solid circles and 
triangles, respectively. 
The RPA results are indicated by horizontal lines in panel (a) 
and by solid squares in panel (b). The open squares in panel (a) 
show the experimental values taken from the compilation 
of Ref. \cite{bnlw}.
}
\label{fig:onep} 
\end{center} 
\end{figure} 

\subsection{Magnetic dipole excitation}

In the $1^+$ excitations, the main part
of the  strength is concentrated in one
excited state, which is dominated by a single
quasi-particle pair. These dominant quasi-particle
configurations are well identified, they are 
the $(1d_{3/2}, 1d_{5/2})_\nu$ for the oxygen isotopes, 
the $(1f_{5/2},1f_{7/2})_\nu$ for the calcium
chain,  and the $(1d_{3/2}, 1d_{5/2})_\pi$ for
the $N=20$ isotones  up to $Z=20$, 
and  the $(1f_{5/2},1f_{7/2})_\pi$ for the heavier ones.
We show in panel (a) of Fig. \ref{fig:onep} the energies 
of the $1^+$ states with the largest $B(M1)$ values
for all the nuclei we have  considered. 
The meaning of the symbols is the same as in Fig. \ref{fig:wpar}. 
The available experimental values, taken from the compilation of 
Ref. \cite{bnlw}, are shown by the open squares.  
The effects of the pairing on these energy values  
are certainly smaller than the differences with the experimental data. 

In the panel (b) of Fig. \ref{fig:onep} we represent the $B(M1)$ values of
these states. We observe that for each group of nuclei they increase
up to a maximum corresponding to the doubly magic nucleus, and then
they decrease. This behavior is related to the
occupation $v^2$ of the states involved. If we consider, for example,
the calcium isotope chain, in the $^{40}$Ca nucleus the two
s.p. states forming the $(1f_{5/2},1f_{7/2})_\nu$ configuration are
empty. The neutron $1f_{7/2}$ state is partially occupied in $^{42}$Ca
and its occupation progressively increases until it reaches full
occupancy in $^{48}$Ca, where $B(M1)$ is maximum.  In all these
isotopes the neutron $1f_{5/2}$ state is empty.  In $^{50}$Ca,
$^{52}$Ca, and $^{54}$Ca the $B(M1)$ values remain almost stable
because, in these isotopes, the additional neutrons occupy the
$2p_{3/2}$ s.p. state and, in the latter nucleus, also the $2p_{1/2}$,
therefore the main neutron configuration is not affected. The $B(M1)$
value reduces for the heavier isotopes since the neutron $1f_{5/2}$ state
begins to be occupied and this diminishes the probability of the
transition.  This trend continues up to $^{60}$Ca where the $1f_{5/2}$
s.p. state is fully occupied and the configuration
$(1f_{5/2},1f_{7/2})_\nu$ is not any more available.

\begin{figure}[!b] 
\begin{center} 
\includegraphics [scale=0.52,angle=0]{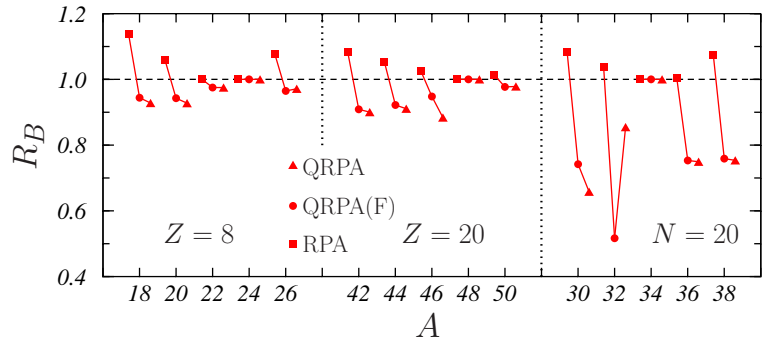} 
\vskip -0.3cm
\caption{\small Values of $\dbval$, 
as defined by Eq. (\ref{eq:dbval}), for those nuclei where in the main
configurations forming the $1^+$ states, i.e.
 $(1d_{3/2},1d_{5/2})_\nu$ and $(1f_{5/2},1f_{7/2})_\nu$ for oxygen 
and calcium isotopes, respectively, and $(1d_{3/2},1d_{5/2})_\pi$ 
for the $N=20$ isotones, one of the s.p. states is partially occupied.
}
\label{fig:v2_1+} 
\end{center} 
\end{figure} 

The effect we have described has a remarkable experimental
evidence in the electron scattering experiments of Ref. \cite{ste80} 
even though, to the best of our knowledge, 
the small strength found in $^{44}$Ca 
with respect to the theoretical expectations remains unexplained. 

We verified the validity of our interpretation of the $B(M1)$ behavior
 by considering the quantity
\beq
\dbval \, = \,  \frac {1}{(\uhfmu)^2\, 
(v_{\mu'}^{\rm HF})^2} \, \frac {B(M1)} {B(M1)_{\rm cs}}
\label{eq:dbval}
\, ,
\eeq
where $B(M1)_{\rm cs}$ indicates the $B(M1)$ value of a closed shell
nucleus in each chain, specifically $^{24}$O, $^{48}$Ca and $^{34}$Si, and
$(\uhfmu)^2$ and $(v_{\mu'}^{\rm HF})^2$ are the HF occupation
probabilities of the quasi-particles involved in the dominant configuration above
mentioned.  In the limiting case where the $1^+$ state is composed
only by the $(\mu \mu')$ quasi-particle configuration, and the pairing
is switched off, $\dbval$ is equal to unity. In this case, the
behavior of the $B(M1)$ values follows that of the corresponding HF
occupation probabilities.

We show in Fig. \ref{fig:v2_1+} the $\dbval$ values for those nuclei
where the configurations we are studying are active.  The results are
located around the unity within a 10\% confirming our interpretation
and indicating that pairing effects are negligible.  The only
exceptions are those of $^{30}$Ne and $^{32}$Mg nuclei. In the former
case the large difference is due to the mixing of the main proton
configuration with the $(3s_{1/2}, 2s_{1/2})_\nu$ component. In the
case of the $^{32}$Mg the main pair is mixed with the $(3s_{1/2},
1d_{3/2})_\nu$ configuration in the QRPA(F) calculations and this
produces a separation of the $1^+$ strength in two peaks. The
inclusion of the $G$ terms partially removes this mixing almost
recovering the RPA result.  In both cases the mixing of the main
configuration with another one is due to small changes of the
s.p. energies which generate s.p. transitions with very similar
energies.

\begin{figure}[!b] 
\begin{center} 
\includegraphics [scale=0.4,angle=0]{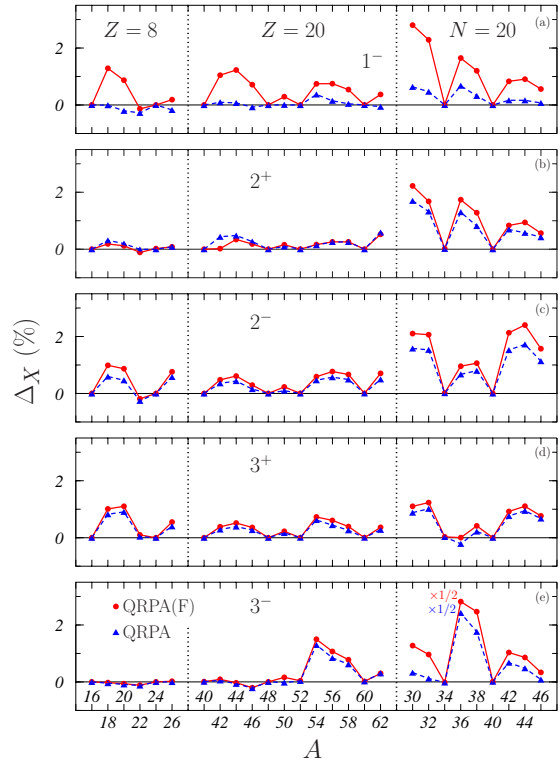} 
\vskip -3mm
\caption{\small 
Relative differences $\Delta_X$ in percentage, 
as defined in Eq. (\ref{eq:diff-centroid}), 
for 1$^-$, $2^\pm$, and $3^\pm$ multipoles 
of the various nuclei investigated. 
Solid circles and triangles indicate QRPA(F) and QRPA results, 
respectively. The values of the $3^-$ excitation in $^{36}$S 
have been divided by 2.}
\label{fig:centr} 
\end{center} 
\end{figure} 

\subsection{Giant resonances}

We have presented so far the effects of the pairing on the low-lying
excitations, and now we discuss the giant resonance region where
the main part of the strength is concentrated. We summarize the large
information regarding this energy region by considering the centroid
energies: 
\beq
\omega_{\rm centr} \, = \, 
\frac {\displaystyle\sum_{\omega=\omega_{\rm min}}^{\omega_{\rm max}} \omega \, B(\omega,0^+ \rightarrow  J^\Pi)} 
      {\displaystyle \sum_{\omega=\omega_{\rm min}}^{\omega_{\rm max}}  \, B(\omega,0^+ \rightarrow  J^\Pi)} 
			\, .
\eeq
Here we have used $\omega_{\rm min}=0$ for all the multipoles with the
exception of the $2^+$, where we have considered $\omega_{\rm min}=10$
MeV to eliminate the large contribution of the low-lying states dominated by 
conjugate configurations. The sums
extend up to $\omega_{\rm max}=80\,$MeV in all cases.  The multipoles
$1^-$, where we excluded the spurious states, $2^\pm$ and $3^\pm$ have
been considered. We did not calculate the centroids for the $1^+$
states because their strengths are concentrated in a single state as
we have already mentioned.
The consistency of our calculations have been tested
by verifying the exhaustion of the Thomas-Reiche-Khun 
and the isoscalar sum rules \cite{har01}.  
We obtained results analogous to those found in RPA 
calculations \cite{don11a,co13}. More details are given in 
Appendix \ref{sec:appoxy}.

In Fig. \ref{fig:centr} we show the relative differences 
\beq
\Delta_{X} \, = \, \displaystyle \frac{\omega^X_{\rm centr}\,-\,\omega_{\rm centr}^{\rm RPA}}{\omega^X_{\rm centr}\,+\,\omega_{\rm centr}^{\rm RPA}}
\, ,
\label{eq:diff-centroid}
\eeq
with $X\equiv \,\,$ \QRPAF \/ and QRPA. The
solid circles indicate the \QRPAF\/ results and
the triangles those of the QRPA calculations.
Except in a few cases, these relative differences are positive,
indicating that the pairing is enhancing the position of the centroid.
In almost all nuclei and multipolarities, the differences are larger in QRPA(F) 
than in QRPA.  The largest effect is seen for the $N=20$
isotones and the largest differences between QRPA(F) and QRPA
calculations are found for the $1^-$ multipolarity.

\section{Summary and conclusions}    
\label{sec:summary} 
We have investigated the excitation spectrum of 6 oxygen isotopes, 12
calcium isotopes and 9 isotones with $N=20$ by using a HF+BCS+QRPA
approach with the aim of studying global effects of the
pairing.  We have used the same effective nucleon-nucleon force 
for the HF part of our calculations, for the pairing sector, and as 
residual interaction in QRPA. We presented
results obtained by using the D1M
parameterization of the Gogny force \cite{ber91}.  In our approach,
pairing effects enter in the construction of the s.p.  configuration
space generated by the HF+BCS calculations, and in the 
results of the QRPA where, in presence of pairing,  the $G$ terms of Eqs. 
(\ref{eq:mata}) and (\ref{eq:matb}) become active. We
carried out calculations that we labelled RPA
by using HF s.p. states and without $G$ terms,
and also calculations  with HF+BCS s.p. states
but without $G$ terms, that we called QRPA(F).  The comparison between
RPA, QRPA(F) and full HF+BCS+QRPA calculations,
that we called QRPA, 
has been used to disentangle the source of the pairing 
effects. We summarize here below the main results of our study.
\begin{itemize}
\item The s.p. wave functions, occupation probabilities  and energies,
  together with the effective nucleon-nucleon interaction are the
  input of the QRPA calculations. The pairing effects on this
  input are related to the modifications of the s.p. energies and
  of the occupation probabilities with respect to the HF values. We
  found a poor correlation between the effects of the pairing on
  these two quantities. In other words, our results show that large
  modifications of the occupation probabilities do not imply large
  modifications of the s.p. energies, and viceversa (see
  Fig. \ref{fig:ve}).
  
\item We observed that, in general, QRPA(F) and QRPA energies are
  larger than those obtained in RPA calculations.  We also found that, for
  each excited state, the QRPA(F) energies are larger than those of
  the QRPA.  The differences between QRPA(F) and QRPA energies are
  smaller than those with the RPA.  We concluded that the pairing
  effect on the s.p. input produces an increase of the excitation
  energies, while in the QRPA calculations it acts in the opposite
  direction. The former effect is larger than that of the other
  one such as the final results is that QRPA energies are still
  larger than those of the RPA (see Figs. \ref{fig:wpar} and
  \ref{fig:dwpar}).

\item While the effects of the pairing on the excitation energy are
  well identified and have a common trend, those concerning the
  changes on the many-body wave functions are more difficult to single
  out and describe.  We have defined an index $\axy$,
  Eq. (\ref{eq:axy}), to quantify the degree of collectivity of an excited
  state. We found general trends only in quadrupole 
  excitations where our results indicate that the pairing increases the
  collectivity of the $2^+$ states while it reduces that of the $2^-$
  states (see Table \ref{tab:axy}).

 \item We have observed that pairing effects on the $1^-$ excitation,
   especially on the PDR, are negligible (see Fig. \ref{fig:pdr}).

\item The study of the first excited $2^+$ states dominated by
  conjugate transitions allows a comparison with other calculations
  and shows that our pairing effects are relatively small (see
  Fig. \ref{fig:other2+}).

\item The strength of the $1^+$ excitation is concentrated on a single
  state, but we did not find remarkable global pairing effects. Specific
  cases deserve a more thorough investigation (see
  Figs. \ref{fig:onep} and \ref{fig:v2_1+}).

\item The study of the excitations in the giant resonance region
  indicates that the energies of the giant resonance centroids increase when
  the pairing is included (see Fig. \ref{fig:centr}).
 
\end{itemize}

The description of the experimental data for the excited states and
nuclei studied does not significantly improve if pairing is taken into account. 
In general, the pairing effects are smaller than the
discrepancy with the experimental data.  We cannot exclude the
possibility that a different parameterization of the effective
nucleon-nucleon interaction, with the same fitting qualities of the
Gogny force, can produce larger pairing effects and a better 
description of the experimental data.  However, we think more
probable that these discrepancies are due to physical phenomena that
the HF+BCS+QRPA approach cannot describe.

\appendix 
\section{Results for $^{18}$O} 
\label{sec:appoxy}
In this appendix we discuss some details regarding the convergence of
our HF+BCS+QRPA calculations.  As example we present the results
obtained for the $^{18}$O nucleus. We conducted analogous tests for
the heavier nuclei we have considered in our investigation and we
obtained similar results.

We carried out HF and BCS calculations by using a box integration
radius of $12\,$fm.  We obtained binding energies of $136.54\,$MeV in
HF and $138.92\,$MeV in HF+BCS, to be compared with the experimental
value of $139.8\,$MeV \cite{bnlw}.  The value obtained by carrying out a
HFB calculation with the approach developed in \cite{ang01a} is $139.8\,$MeV.

\begin{table}[!t] 
\begin{center} 
\begin{tabular}{cc r ccrr ccrr} 
\hline \hline 
  &~~& HF &~~& \multicolumn{3}{c}{HF+BCS} &~~& \multicolumn{3}{c}{HFB} \\
\cline{3-3}\cline{5-7}\cline{9-11}
s.p. state & &\multicolumn{1}{c}{$\epsilon_\mu$} & & $v_\mu^{2}$ &  
\multicolumn{1}{c}{$E_\mu$} & \multicolumn{1}{c}{$\bcse_\mu$}
 & &$v_\mu^{2}$ &  
\multicolumn{1}{c}{$E_\mu$} & \multicolumn{1}{c}{$\epsilon_\mu$} \\
\hline
 $1s_{1/2}$ && -36.73 && 0.999 & 30.35 & -36.78 && 0.998 & 30.37 & -36.88  \\
 $1p_{3/2}$ && -19.95 && 0.995 & 13.65 & -20.08 && 0.991 & 14.31 & -20.66 \\
 $1p_{1/2}$ && -15.79 && 0.991 &  9.53 & -15.96 && 0.983 &  9.39 & -15.72 \\
 $1d_{5/2}$ &&  -5.88 && 0.332 &  1.63 &  -4.80 && 0.323 & 2.25 &  -5.81\\
 $2s_{1/2}$ &&  -2.68 && 0.010 &  3.83 &  -2.60 && 0.025 & 4.28 &  -1.98\\
 $1d_{3/2}$ &&   1.16 && 0.004 &  7.65 &   1.22 && 0.011 & 8.24 &  2.77 \\
\hline\hline 
\end{tabular} 
\caption{\small Occupation probabilities and energies of the 
neutron quasi-particle states obtained in HF,
HF+BCS and HFB calculations for neutron states in 
$^{18}$O with $v^2 < 10^{-4}$.
The HF energies $\epsilon_\mu$, the quasi-particle energies 
$E_\mu$, defined in Eq. (\ref{eq:qpenergy}), 
and the energies $\bcse_\mu$, defined in  
Eq. (\ref{eq:bcse}), are expressed in MeV.  
} 
\label{tab:nstate} 
\end{center} 
\end{table} 

In Table \ref{tab:nstate} we give the HF energies $\epsilon_\mu$, the
occupation probabilities $v_\mu^{2}$, the quasi particle energies
$E_\mu$, and the BCS s.p. energies $\bcse_\mu$ for the neutron
s.p. states with $v_\mu^{2}>10^{-4}$. For sake of completeness we give
the values of the same quantities obtained in a HFB calculation.

The configuration space used in the QRPA calculations includes all the
s.p. states with energy smaller than a chosen value $E_p$. 
We
  insert a further restriction regarding the number of quasi-particle
  pairs. We neglect those pairs composed by quasi-particles whose
  energy is, individually, larger than $E_{pp}$.  This second
  restriction is useful in heavy nuclei, but, in the case of the nuclei
studied in the present work, we have always used
$E_{p}=E_{pp}$. Nevertheless, the use of these two cutoff energies is
not enough to generate QRPA matrices of dimension smaller than 4000,
which is our numerical limitation.  The problem is generated by the
large number of configuration pairs where both s.p. states have
$\epsilon_\mu > \lambda$.  For the excited states of our interest,
i.e. the low-lying ones and the giant resonances, the contribution of
many of these last type of pairs, is scarcely relevant.  These are
pairs where the wave functions describing both quasi-particle states
are non-resonant and lying high in the continuum. For this reason
we further reduce the number of
this type of pairs by omitting those where the occupation probability
$v^2$ of both states forming the pair is smaller than a critical value
$v^2_{\rm cut}$.

%
\begin{table}[!b] 
\begin{center} 
\begin{tabular}{cccccccccc} 
\hline \hline 
~~&~~&~~&~~&~~&\multicolumn{2}{c}{$1^-_1$}& ~~&\multicolumn{2}{c}{$1^-_2$} \\\cline{6-7}\cline{9-10}
calculation &$v^2_{\rm cut}$&& \# of pairs&& $\omega$ (MeV) &  $B(E1)$ $(e^2\,{\rm fm}^2)$ && $\omega$ (MeV) & $B(E1)$ $(e^2\,{\rm fm}^2)$\\
\hline 
A    & - & & 1112 && 2.73 & $1.87 \cdot 10^{-3}$ && 10.62 & $8.18 \cdot 10^{-2}$\\
B   & $10^{-5}$ & &  271 && 2.76 & $1.86 \cdot 10^{-3}$ && 10.64 & $7.73\cdot 10^{-2}$\\
C  & $10^{-3}$ &&   164 && 2.79 & $1.53 \cdot 10^{-3}$ && 10.65 & $7.53 \cdot 10^{-2}$ \\
\hline\hline 
\end{tabular} 
\caption{\small 
Excitation energies and $B(E1)$ values of the first two 1$^-$ states of $^{18}$O 
in the HF+BCS+QRPA approach. In all the cases we used $E_p=E_{pp}=100\,$MeV. 
The results of calculation A have been obtained without any further
limitation of the s.p. configuration space. In calculations B and C 
we have excluded those pairs where both
quasi-particle states have $v_\mu^2 < 10^{-5}$ and
$10^{-3}$, respectively. 
}
\label{tab:conv} 
\end{center} 
\end{table} 

As example, we show in Table \ref{tab:conv} the excitation energies,
and the corresponding $B(E1)$ values, of the first two 1$^-$ excited
states.  The $1^-_1$ state is the spurious isoscalar excitation
related to the centre of mass motion. All the calculations have been
carried out with $E_p=E_{pp}=100\,$MeV. The results of calculation A
have been obtained without any further limitations of the
s.p. configuration space, by handling 1112 pairs.  We reduce this
number by using $v^2_{\rm cut}=10^{-5}$ (calculation B), and
$v^2_{\rm cut}=10^{-3}$ (calculation C).  Even though there is a
remarkable reduction of the number of pairs when the selection related
to $v^2_{\rm cut}$ is activated, the energy eigenvalues are only
modified by a few tens of keV. The $B(E1)$ values are more sensitive
to this reduction.

%
\begin{table}[!t] 
\begin{center} 
\begin{tabular}{c ccc ccc ccc} 
\hline \hline 
  $E_p=E_{pp}$ && $\omega_1$ & $B_1(E1)$ &&  $\omega_2$ & $B_2(E1)$ &&  $\omega_3$ & $B_3(E1)$  \\ 
\hline
    60.0 &  &    3.22 &  $2.98 \cdot 10^{-3}$ &  & 11.16 &  $1.19 \cdot 10^{-2}$ && 11.99 & $1.27 \cdot 10^{-1}$ \\
    80.0 &   &   3.11 &  $2.51 \cdot 10^{-3}$ &  & 11.12 &  $1.00 \cdot 10^{-2}$ &&   11.98 & $1.29 \cdot 10^{-1}$ \\
   100.0 &  &    2.68 & $2.76 \cdot 10^{-3}$ &  & 11.02 &  $7.43 \cdot 10^{-3}$ &&   11.96 & $1.32 \cdot 10^{-1}$ \\
   120.0 &  &    2.55 & $2.64 \cdot 10^{-3}$ &  & 10.98 &  $6.37 \cdot 10^{-3}$ &&   11.95 & $1.33 \cdot 10^{-1}$ \\
   140.0 &  &    2.41 & $1.89 \cdot 10^{-3}$ &  & 10.96 &  $5.85 \cdot 10^{-3}$ &&   11.94 & $1.35 \cdot 10^{-1}$\\
   160.0 &  &    2.16 & $1.71 \cdot 10^{-3}$ &  & 10.94 &  $5.14 \cdot 10^{-3}$ &&   11.94 &  $1.36 \cdot 10^{-1}$ \\
   180.0 &  &    2.12 & $1.62 \cdot 10^{-3}$ &  & 10.93 &  $4.94 \cdot 10^{-3}$ &&   11.93 & $1.36 \cdot 10^{-1}$\\
   200.0 &  &    1.92 & $1.69 \cdot 10^{-3}$ &  & 10.92 &  $4.63 \cdot 10^{-3}$ &&   11.93 & $1.36 \cdot 10^{-1}$ \\
   250.0 &  &    1.87 & $1.34 \cdot 10^{-3}$ & &  10.91 &  $4.48 \cdot 10^{-3}$ &&   11.93 & $1.37 \cdot 10^{-1}$\\
\hline
\hline
\end{tabular} 
\caption{\small 
Excitation energies, $\omega$ in MeV, and
    $B(E1)$ values, in $e^2$ fm$^2$, of the first three $1^-$ excited states 
as a function of the cutoff energies $E_p$ and $E_{pp}$, also expressed in MeV.
}  
\label{tab:conv1-} 
\end{center} 
\end{table} 

We tested the sensitivity of the energy of the spurious isoscalar
$1^-$ state to the dimensions of the configuration space. In Table
\ref{tab:conv1-} we show energies and $B$-values of the first three
$1^-$ states obtained with $v^2_{\rm cut}=10^{-5}$.  We have carried
out calculations by using different values of $E_p=E_{pp}$. It is
evident that the spurious state is extremely sensitive to the size of
the configuration space, while the energies of the other two states
change very little.

We also checked the presence of a low-lying $0^+$ spurious state
generated by the non conservation of the particle number in the
nuclear wave function. Also in this case the energy of this spurious
state is extremely sensitive to the dimensions of the quasi-particle
configuration space. When we used $E_p=E_{pp}=60\,$MeV we found this
$0^+$ state at $0.11\,$MeV. With larger configuration spaces,
$E_p=E_{pp} > 100\,$MeV, it is below zero and the QRPA secular
equations (\ref{eq:QRPA}) have an imaginary solution.

We tested the stability of our solutions by changing of 10\% the $E_p$
and $E_{pp}$ values. In all cases we obtained differences in the QRPA
energies within the numerical accuracy.

The results presented in this paper have been carried out by using the
values $E_p=E_{pp}=200\,$MeV and $v^2_{\rm cut}=10^{-3}$, which ensure
the stability of the results for all the nuclei studied, even
in $^{62}$Ca, the heaviest one.

Finally, we evaluated energy weighted sum rules by integrating up to
the excitation energy of 50 MeV, and using $E_p=E_{pp}=200\,$MeV. For
the $1^-$ excitation we calculated the Thomas-Reiche-Khun sum rule
which we found exhausted up to the 0.95 of the expected value, this
last one calculated by considering an enhancement factor of $\kappa =
0.5$ with respect to the traditional value \cite{nak09,don11a}).  For
the isoscalar $0^+$ and $1^-$ excitations \cite{har01,ter05} we found
an over estimation of the expected values of about 1.1, to be compared
with the values of 0.86 and 0.96 obtained, respectively, when RPA
calculations have been carried out. The source of this discrepancy is
under investigation.

\acknowledgments  
This work has been partially supported by  
the Junta de Andaluc\'{\i}a (FQM387), the Spanish Ministerio de 
Econom\'{\i}a y Competitividad (FPA2015-67694-P) and the European 
Regional Development Fund (ERDF). 
 

\begin{thebibliography}{10}
\expandafter\ifx\csname url\endcsname\relax
  \def\url#1{\texttt{#1}}\fi
\expandafter\ifx\csname urlprefix\endcsname\relax\def\urlprefix{URL }\fi
\expandafter\ifx\csname href\endcsname\relax
  \def\href#1#2{#2} \def\path#1{#1}\fi

\bibitem{boh58}
A.~Bohr, B.~Mottelson, D.~Pines, Phys. Rev. 110 (1958) 936.

\bibitem{bar60}
M.~Baranger, Phys. \ Rev. 120 (1960) 957.

\bibitem{ang14}
M.~Anguiano, A.~M. Lallena, G.~Co', V.~{De Donno}, J. \ Phys. \ G 41 (2014)
  025102.

\bibitem{ang15}
M.~Anguiano, A.~M. Lallena, G.~Co', V.~{De Donno}, J. \ Phys. \ G 42 (2015)
  079501.

\bibitem{ang16a}
M.~Anguiano, R.~N. Bernard, A.~M. Lallena, G.~Co', V.~{De Donno}, Nucl. \ Phys.
  \ A 955 (2016) 181.

\bibitem{dec80}
J.~Decharg\'e, D.~Gogny, Phys. \ Rev. \ C 21 (1980) 1568.

\bibitem{paa07}
N.~Paar, D.~Vretenar, E.~Khan, G.~Col\`o, Rep. Prog. Phys. 70 (2007) 691.

\bibitem{kha02}
E.~Khan, N.~Sandulescu, M.~Grasso, N.~{Van Giai}, Phys. \ Rev. \ C 66 (2002)
  024309.

\bibitem{yam04}
M.~Yamagami, N.~{Van Giai}, Phys. \ Rev. \ C 69 (2004) 034301.

\bibitem{ter05}
J.~Terasaki, J.~Engel, M.~Bender, J.~Dobaczewski, W.~Nazarewicz, M.~Stoitsov,
  Phys. \ Rev. \ C 71 (2005) 034310.

\bibitem{car12}
B.~G. Carlson, J.~Toivanen, A.~Pastore, Phys. \ Rev. \ C 86 (2012) 014307.

\bibitem{yuk14}
E.~Y{\"u}ksel, N.~{Van Giai}, E.~Khan, K.~Bozkurt, Phys. \ Rev. \ C 89 (2014)
  064322.

\bibitem{gia03}
G.~Giambrone, S.~Scheit, F.~Barranco, P.~Bortignon, G.~Col{\`o}, D.~Sarchi,
  E.~Vigezzi, Nucl. \ Phys. \ A 726 (2005) 3.

\bibitem{mar11}
M.~Martini, S.~P\'eru, M.~Dupuis, Phys. \ Rev. \ C 83 (2011) 034309.

\bibitem{paa03}
N.~Paar, P.~Ring, T.~Nik\v{s}i\'c, D.~Vretenar, Phys. \ Rev. \ C 67 (2003)
  034312.

\bibitem{cao05}
L.-G. Cao, Z.-Y. Ma, Phys. \ Rev. \ C 71 (2005) 034305.

\bibitem{gor09}
S.~Goriely, S.~Hilaire, M.~Girod, S.~P\'eru, Phys. \ Rev. \ Lett. 102 (2009)
  242501.

\bibitem{co98b}
G.~Co', A.~M. Lallena, Nuovo \ Cimento \ A 111 (1998) 527.

\bibitem{rin80}
P.~Ring, P.~Schuck, The nuclear many-body problem, Springer, Berlin, 1980.

\bibitem{suh07}
J.~Suhonen, From nucleons to nucleus, Springer, Berlin, 2007.

\bibitem{edm57}
A.~R. Edmonds, Angular momentum in quantum mechanics, Princeton University
  Press, Princeton, 1957.

\bibitem{cat94}
F.~Catara, N.~{Dinh Dang}, M.~Sambataro, Nucl. \ Phys. \ A 579 (1994) 1.

\bibitem{sch96}
J.~Schwieger, F.~Simkovic, A.~Faessler, Nucl. \ Phys. \ A 600 (1996) 179.

\bibitem{mar98}
A.~Mariano, J.~G. Hirsch, Phys. \ Rev. \ C 57 (1998) 3015.

\bibitem{cha07t}
F.~Chappert, Nouvelles param\'etrisation de l'interaction nucl\'eaire effective
  de {Gogny}, Ph.D. thesis, Universit\'e de Paris-Sud XI (France),
  http://tel.archives-ouvertes.fr/tel-001777379/en/ (2007).

\bibitem{ber91}
J.~F. Berger, M.~Girod, D.~Gogny, Comp. \ Phys. \ Commun. 63 (1991) 365.

\bibitem{egi95}
J.~L. Egido, J.~Lessing, V.~Martin, L.~M. Robledo, Nucl. \ Phys. \ A 594 (1995)
  70.

\bibitem{don14a}
V.~{De Donno}, G.~Co', M.~Anguiano, A.~M. Lallena, Phys. \ Rev. \ C 89 (2014)
  014309.

\bibitem{bau99}
A.~R. Bautista, G.~Co', A.~M. Lallena, Nuovo \ Cimento \ A 112 (1999) 1117.

\bibitem{ang11}
M.~Anguiano, G.~Co', V.~{De Donno}, A.~M. Lallena, Phys. \ Rev. \ C 83 (2011)
  064306.

\bibitem{tol11}
S.~V. Tolokonnikov, S.~Kamerdzhiev, D.~Voitenkov, S.~Krewald, E.~E. Saperstein,
  Phys. \ Rev. \ C 84 (2011) 064324.

\bibitem{don11a}
V.~{De Donno}, G.~Co', M.~Anguiano, A.~M. Lallena, Phys. \ Rev. \ C 83 (2011)
  044324.

\bibitem{mat05}
M.~Matsuo, K.~Mizuyama, Y.Serizawa, Phys. \ Rev. \ C 71 (2005) 064326.

\bibitem{bnlw}
{Brookhaven National Laboratory}, National nuclear data center,
  http://www.nndc.bnl.gov/.

\bibitem{co13}
G.~Co', V.~{De Donno}, M.~Anguiano, A.~M. Lallena, Phys. Rev. C 87 (2013)
  034305.

\bibitem{ram01}
S.~Raman, C.~W. {Nestor Jr.}, P.~Tikkanen, Atomic \ Data \ and \ Nuclear \ Data
  \ Tables 78 (2001) 1.

\bibitem{cha98}
E.~Chabanat, P.~Bonche, P.~Haensel, J.~Meyer, F.~Schaeffer, Nucl. \ Phys. \ A
  635 (1998) 231.

\bibitem{bar82}
J.~Bartel, P.~Quentin, M.~Brack, C.~Guet, H.~B. H{\aa}kansson, Nucl. \ Phys. \
  A 386 (1982) 79.

\bibitem{ste80}
W.~Steffen, H.-D. Gr{\"a}f, W.~Gross, D.~Meuer, A.~Richter, E.~Spamer,
  O.~Titze, W.~Kn{\"u}pfer, Phys. \ Lett. \ B 95 (1980) 23.

\bibitem{har01}
M.~N. Harakeh, A.~van~der Woude, Giant resonances, Claredon press, Oxford,
  2001.

\bibitem{ang01a}
M.~Anguiano, J.~L. Egido, L.~M. Robledo, Nucl. \ Phys. \ A 683 (2001) 227.

\bibitem{nak09}
H.~Nakada, K.~Mizuyama, M.~Yamagami, M.~Matsuo, Nucl. \ Phys. A 828 (2009) 283.

\end{thebibliography}

%
 
\end{document}